\documentclass[twocolumn]{aastex631}

\usepackage{natbib}           % Bibliographic stuff
\usepackage{graphicx}	      % Including figure files
\usepackage{amsmath}	      % Advanced maths commands - this affect the appendix section! 
\usepackage{amssymb}	      % Extra maths symbols
\usepackage{multirow}

\usepackage{booktabs}
\usepackage{threeparttable}
\usepackage{placeins}

\newcommand{\mum}{\ifmmode{\rm \mu m}\else{$\mu$m }\fi}             % micron symbol
\newcommand{\Zsun}{$Z_{\odot}$}		                	    	    % solar metallicity
\providecommand{\e}[1]{\ensuremath{\times 10^{#1}}}	            % for scientific notation
\newcommand{\chisq}{\ifmmode{\chi^{2} }\else{$\chi^2$}\fi}
\newcommand{\rchisq}{\ifmmode{\chi^{2} }\else{$\chi^2_\nu$}\fi}

\newcommand{\hst}{{\em HST}}               	         		% for HST

\newcommand{\Hii}{H{\sc ii} }                               % for H II regions
\newcommand{\h}{H$_{2}$}
\newcommand{\paa}{Pa-$\alpha$} %1.8751 \microns 

\shorttitle{{\it JWST} NGC 346 Imaging}   %% Give here short title %%
\graphicspath{{./}{figures/}}              %% Set Figure path %%
\begin{document}

\title{Young Stellar Objects in NGC 346: A JWST NIRCam/MIRI Imaging Survey}

\author[0000-0002-2667-1676]{Nolan Habel}
\affil{Jet Propulsion Laboratory, California Institute of Technology, 4800 Oak Grove Dr., Pasadena, CA 91109, USA}
\correspondingauthor{Nolan Habel}
\email{nolan.m.habel@jpl.nasa.gov}

\author[0000-0002-7512-1662]{Conor Nally}
\affil{Institute for Astronomy, University of Edinburgh, Royal Observatory, Blackford Hill, Edinburgh EH9 3HJ, UK}

\author[0000-0003-4023-8657]{Laura Lenki\'{c}}
\affil{Jet Propulsion Laboratory, California Institute of Technology, 4800 Oak Grove Dr., Pasadena, CA 91109, USA}
\affil{Stratospheric Observatory for Infrared Astronomy, NASA Ames Research Center, Mail Stop 204-14, Moffett Field, CA 94035, USA}

\author[0000-0002-0522-3743]{Margaret Meixner}
\affil{Jet Propulsion Laboratory, California Institute of Technology, 4800 Oak Grove Dr., Pasadena, CA 91109, USA}
\affil{Stratospheric Observatory for Infrared Astronomy, NASA Ames Research Center, Mail Stop 204-14, Moffett Field, CA 94035, USA}

\author[0000-0001-7906-3829]{Guido De Marchi}
\affil{European Space Research and Technology Centre, Keplerlaan 1, 2200 AG Noordwijk, The Netherlands}

\author[0000-0001-6872-2358]{Patrick \ J.\ Kavanagh}
\affil{Department of Experimental Physics, Maynooth University, Maynooth, Co. Kildare, Ireland}

\author[0000-0003-2902-8608]{Katja Fahrion}
\affil{European Space Research and Technology Centre, Keplerlaan 1, 2200 AG Noordwijk, The Netherlands}

\author[0000-0001-6576-6339]{Omnarayani Nayak}
\affil{Space Telescope Science Institute, 3700 San Martin Drive, Baltimore, MD 21218, USA}
\affil{NASA Goddard Space Flight Center, 8800 Greenbelt Road, Greenbelt, MD, USA}

\author[0000-0002-2954-8622]{Alec S.\ Hirschauer}
\affil{Space Telescope Science Institute, 3700 San Martin Drive, Baltimore, MD 21218, USA}

% Add names in alphabetical order here: 

%\author{A. N. Other}
%\affil{Space Telescope Science Institute, 3700 San Martin Drive, Baltimore, MD 21218, USA}

\author[0000-0003-4870-5547]{Olivia C.\ Jones}
\affil{UK Astronomy Technology Centre, Royal Observatory, Blackford Hill, Edinburgh, EH9 3HJ, UK}

\author[0000-0002-1892-2180]{Katia Biazzo}
\affil{1INAF, Astronomical Observatory of Rome, Via Frascati 33,
Monteporzio Catone, I-00078, Italy}

\author[0000-0001-9737-169X]{Bernhard R. Brandl}
\affil{Leiden Observatory, Leiden University,
PO Box 9513, 2300 RA Leiden, The Netherlands}

\author[0000-0002-0577-1950]{J. Jaspers}
\affil{Dublin Institute for Advanced Studies, School of Cosmic Physics, Astronomy \& Astrophysics Section 31 Fitzwilliam Place, Dublin 2, Ireland}
\affil{Department of Experimental Physics, Maynooth University, Maynooth, Co. Kildare, Ireland}

\author[0000-0002-0786-7307]{Klaus M. Pontoppidan}
\affil{Space Telescope Science Institute, 3700 San Martin Drive, Baltimore, MD 21218, USA}

\author[0000-0002-9573-3199]{Massimo Robberto}
\affil{Space Telescope Science Institute, 3700 San Martin Drive, Baltimore, MD 21218, USA}
\affil{Department of Physics \& Astronomy, Johns Hopkins University, 3400 N.\ Charles St., Baltimore, MD 21218, USA}

\author[0000-0001-5742-2261]{C. Rogers}
\affil{Leiden Observatory, Leiden University,
PO Box 9513, 2300 RA Leiden, The Netherlands}

\author[0000-0003-2954-7643]{E.\ Sabbi}
\affil{Space Telescope Science Institute, 3700 San Martin Drive, Baltimore, MD 21218, USA}

\author[0000-0001-9855-8261]{B.\ A.\ Sargent}
\affil{Space Telescope Science Institute, 3700 San Martin Drive, Baltimore, MD 21218, USA}
\affil{Department of Physics \& Astronomy, Johns Hopkins University, 3400 N.\ Charles St., Baltimore, MD 21218, USA}

\author[0000-0002-0322-8161]{David R. Soderblom} 
\affil{Space Telescope Science Institute, 3700 San Martin Drive, Baltimore, MD 21218, USA}

\author[0000-0002-6091-7924]{Peter\ Zeidler} 
\affil{AURA for the European Space Agency,
Space Telescope Science Institute, 3700 San Martin Drive, Baltimore, MD 21218, USA}

%% Note that the \and command from previous versions of AASTeX is now
%% depreciated in this version as it is no longer necessary. AASTeX 
%% automatically takes care of all commas and "and"s between authors names.

%% AASTeX 6.31 has the new \collaboration and \nocollaboration commands to
%% provide the collaboration status of a group of authors. These commands 
%% can be used either before or after the list of corresponding authors. The
%% argument for \collaboration is the collaboration identifier. Authors are
%% encouraged to surround collaboration identifiers with ()s. The 
%% \nocollaboration command takes no argument and exists to indicate that
%% the nearby authors are not part of surrounding collaborations.

%% Mark off the abstract in the ``abstract'' environment. 
\begin{abstract}

We present a {\em JWST} ~imaging survey with NIRCam and MIRI of NGC 346, the brightest star-forming region in the Small Magellanic Cloud (SMC). By combining aperture and point spread function (PSF) photometry of eleven wavelength bands across these two instruments, we have detected more than 200,000 unique sources. Using near-infrared (IR) color analysis, we observe various evolved and young populations, including 196 young stellar objects (YSOs) and pre-main sequence stars suitable for forthcoming spectroscopic studies. We expand upon this work, creating mid-IR color-magnitude diagrams and determining color cuts to identify 833 reddened sources which are YSO candidates. We observe that these candidate sources are spatially associated with regions of dusty, filamentary nebulosity. Furthermore, we fit model YSO spectral energy distributions (SEDs) to a selection of sources with detections across all of our MIRI bands. We classify with a high degree of confidence 23 YSOs in this sample and estimate their radii, bolometric temperatures, luminosities, and masses. We detect YSOs approaching 1 M$_{\odot}$, the lowest-mass extragalactic YSOs confirmed to date.

\end{abstract}

%% Keywords should appear after the \end{abstract} command. 
%% The AAS Journals now uses Unified Astronomy Thesaurus concepts:
%% https://astrothesaurus.org
%% You will be asked to selected these concepts during the submission process
%% but this old "keyword" functionality is maintained in case authors want
%% to include these concepts in their preprints.

%% From the front matter, we move on to the body of the paper.
%% Sections are demarcated by \section and \subsection, respectively.
%% Observe the use of the LaTeX \label
%% command after the \subsection to give a symbolic KEY to the
%% subsection for cross-referencing in a \ref command.
%% You can use LaTeX's \ref and \label commands to keep track of
%% cross-references to sections, equations, tables, and figures.
%% That way, if you change the order of any elements, LaTeX will
%% automatically renumber them.
%%
%% We recommend that authors also use the natbib \citep
%% and \citet commands to identify citations.  The citations are
%% tied to the reference list via symbolic KEYs. The KEY corresponds
%% to the KEY in the \bibitem in the reference list below. 

\section{Introduction} 
\label{sec:intro}

Residing in the Small Magellanic Cloud (SMC), NGC 346 is a region of intense and active star formation. This young cluster is the brightest star-forming region in this relatively nearby ($\sim$62~kpc; \citealt{deGrijs2015}) metal-poor dwarf galaxy ($\sim$1/5~\Zsun; \citealt{Peimbert2000}). It is a prime laboratory to study low-metallicity star formation, and with its environment resembling that of the early universe at the epoch of peak star formation \citep[$z \sim 2$;][]{Dimaratos2015} it offers the ability to study these processes at a sub-parsec resolution. We summarize the properties of NGC 346 in \autoref{tab:NGC346_properties}.

NGC 346 has a complex star formation history, with multiple populations identified within it from massive, evolved stars, to low-mass young stellar objects (YSOs) \citep[e.g.,][]{Cignoni2011}.
Thirty massive (35--100 M$_\odot$) O-type stars drive NGC 346's H\,{\sc ii} region, powering its radiative and mechanical feedback, enriching the interstellar environment \citep{Massey1989, Evans2006}.
Past optical and infrared (IR) studies have revealed young populations existing throughout this region. Thousands of pre-main sequence (PMS) stars scattered throughout NGC 346's dusty filaments were discovered by \citep{Nota2006} with {\em Hubble Space Telescope} ({\em HST}) imaging, some with masses as low as 0.6--3 M$_\odot$ \citep{Sabbi2007, Hennekemper2008, DeMarchi2011}. 
Surveys of the SMC in the IR with {\em Spitzer} and {\em Herschel} \citep{Bolatto2007, Gordon2011, Meixner2013} discovered approximately 100 candidate YSOs within NGC 346 at early stages in their evolution. These candidates possessed masses as low as 1.5 M$_\odot$ and are estimated to have formed within the last $\sim$1Myr \citep{Simon2007, bib:sewilo2013, Seale2014}. From these sources, \cite{Simon2007} estimated a star formation rate in NGC 346 of $>3.2 \times 10^{-3}$ ${\rm M}_{\odot} \, {\rm yr}^{-1}$.

The NGC 346 complex encompasses a range of stellar and interstellar environments \citep{Hony2015}. Stellar catalog of the region show  multiple linked clusters of varying age and complexity, and an extended hierarchical distribution of stars dispersed across the field \citep{Sabbi2008, Hennekemper2008, Gouliermis2014}. Age estimates for the youngest populations in the extend from $0.01-0.05$Myr \citep{bib:Rubio2018} to $1-3$Myr \citep{Bouret2003,Sabbi2007,DeMarchi2011,bib:Dufton2019}.
This stratification is due to both turbulent star formation and early dynamical evolution \citep{Sabbi2022, Zeidler2022}. 
The interstellar medium (ISM) in NGC 346 is also complex. There is a wide range of surface brightnesses exhibited in polycyclic aromatic hydrocarbon emission (PAH; 8 $\mu$m), warm dust (24 $\mu$m) and molecular gas \citep[CO J = 2--1;][]{Rubio2000, Contursi2000, Hony2015}. 
This emission is closely correlated with that of H$\alpha$, which appears as a well-defined central bar extending from the center of the region to the northeast with an arc-like structure extending from southeast to northwest. 
ALMA CO(J = 1--0) data have revealed clumpy filaments with multiple velocity components. At the intersection of these filaments, clusters of YSOs and young PMS star clusters are detected, formed as a result of a cloud-cloud collision 0.2 Myr ago \citep{Neelamkodan2021}.

The 6.5m {\em JWST}, launched in December 2021, provides for the first time the point-source sensitivity and high spatial resolution in the 1--28 $\mu$m wavelength range to detect and spatially resolve both sub-solar mass YSOs and PMS stars with planet-forming dust disks outside our own Galaxy.
Initial results from NIRCam observations of NGC 346 were presented by \citet{bib:jones2023} and revealed solar-mass YSOs.  This paper builds upon this work with more refined data processing and source extraction for the NIRCam data and introduces the addition of MIRI observations, extending the study of this region into the mid-IR.

As protostars evolve, their envelope falls onto a circumstellar disk, or is reprocessed into the ISM by feedback, becoming progressively more optically thin and revealing the central protostar and disk \citep{bib:robitaille06}. For the youngest sources, the dusty envelope and disk re-radiate stellar light at mid-IR wavelengths. With the evolution of the protostar and the dispersal or thinning of the envelope, YSOs can be observed at shorter, near-IR wavelengths. Past surveys of star forming regions in the Milky Way with {\em HST}, such as in the Orion Molecular cloud, have shown that up to a third of YSOs seen in the mid-IR are not detected at near-IR wavelengths \citep{bib:habel21}. Mid-IR observations, such as those introduced in this work, are thus crucial for obtaining a comprehensive census of YSO populations, and particularly important for detecting the youngest YSOs.

%------------------------------------------------------------------------------------
In this paper, we present {\em JWST} imaging data of NGC 346 from 1.15 -- 25.0~$\mu$m and provide an overview of the populations in the region, particularly the young populations, as determined from photometric analysis.
In \autoref{sec:observations} we  discuss the choice of {\em JWST} filters and describe the observations and data reduction. 
We describe the processing of our data and present high-resolution IR images in \autoref{sec:processing}; we examine the general morphology and make comparisons between observations at various wavelengths.
In \autoref{sec:photometry}, we detail our methods for extracting aperture and point spread function (PSF) photometry from our observations and our construction of a source catalog.
In \autoref{sec:results} we present color-magnitude diagram (CMD) and color-color diagram (CCD) analyses of the populations detected in NGC 346, in particular identifying with confidence $\sim$ 200 YSOs by their near-IR color. We further examine the mid-IR properties of these populations, identifying $\sim$800 sources with IR excess associated with YSOs. Lastly, we combine our near-IR and mid-IR observations, fitting model YSO spectral energy distributions (SEDs) to 23 sources detected in all mid-IR bands. From these fits, we estimate the radii, bolometric temperatures, luminosities, and masses of these objects. The least massive of these YSOs possesses a mass of 0.95 $\mathrm{M_{\odot}}$, the lowest mass extragalactic YSO confirmed to date.
Finally, we summarize our findings in \autoref{sec:conclusion}.

\begin{table}
{\centering
\begin{tabular}{lr}
\hline
\hline
NGC 346 Properties & \\
\hline
\hline
Distance, (kpc) & 62$\pm$1   \citep{deGrijs2015} \\
Metallicity, ($Z_\odot$) & $0.2$$\pm$$0.06$   \citep{Russell1992} \\
$A_{\rm V}$ (mag)& 0.12   \citep{Schlegel1998}  \\
$E_{\rm B-V}$ (mag) & 0.04  \citep{Harris2004}  \\
Gas/dust ratio ($r_{\rm gd}$) & 1250    \citep{Hony2015} \\
\hline
\hline
\end{tabular}
\caption{Various properties of NGC 346 referenced in this work. Extinction laws in \cite{Schlegel1998} are from \cite{bib:cardelli1989} and \cite{bib:odonnell1994}.}
\label{tab:NGC346_properties}

}
\end{table}
%%%%%%%%%%%%%%%%%%%%%%%%%%%%%%%%%%%%%%%%%%

\section{Observations}
\label{sec:observations}

\begin{figure*}[ht]
\begin{minipage}[b]{0.47\linewidth}
\centering
\includegraphics[width=\textwidth]{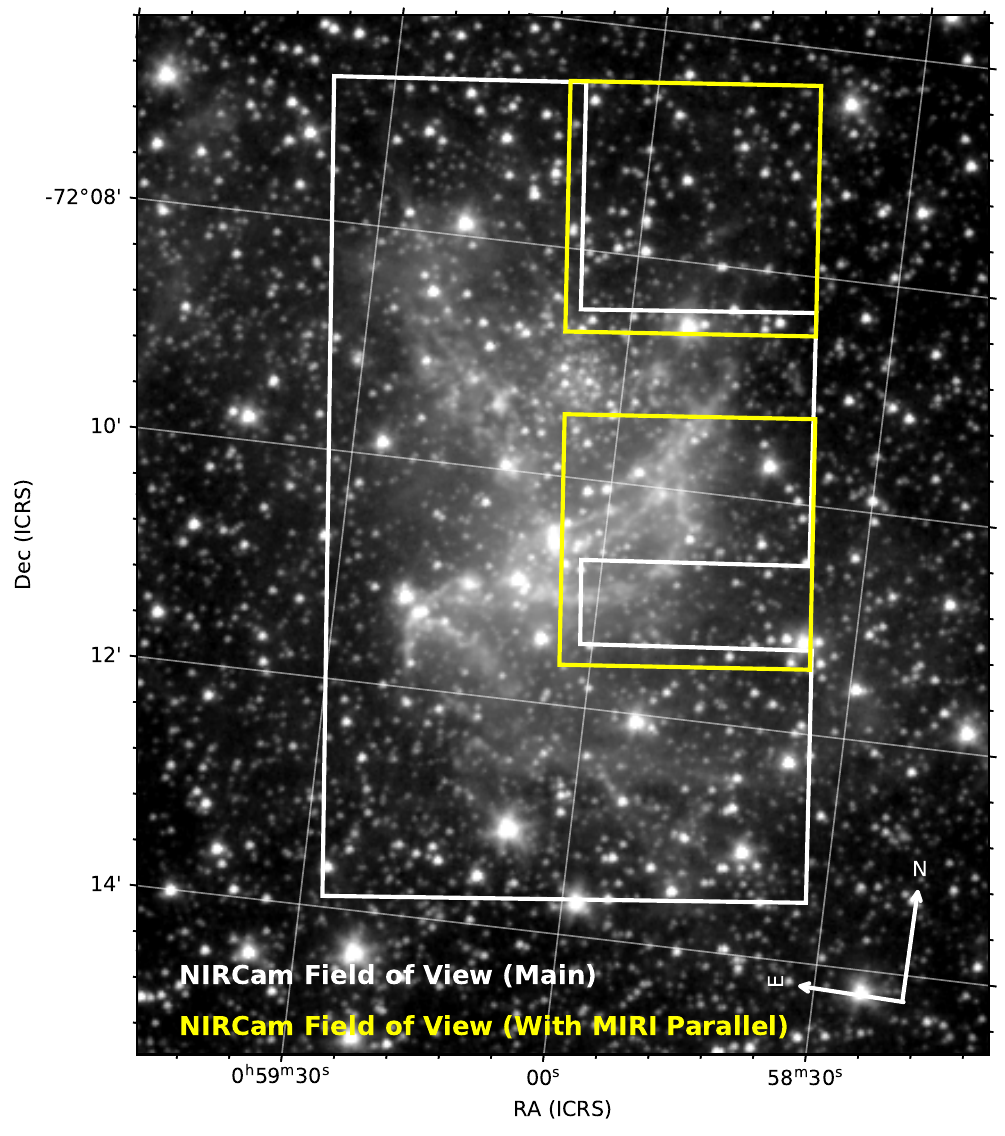}
\end{minipage}
\hspace{0.5cm}
\begin{minipage}[b]{0.47\linewidth}
\centering
\includegraphics[width=\textwidth]{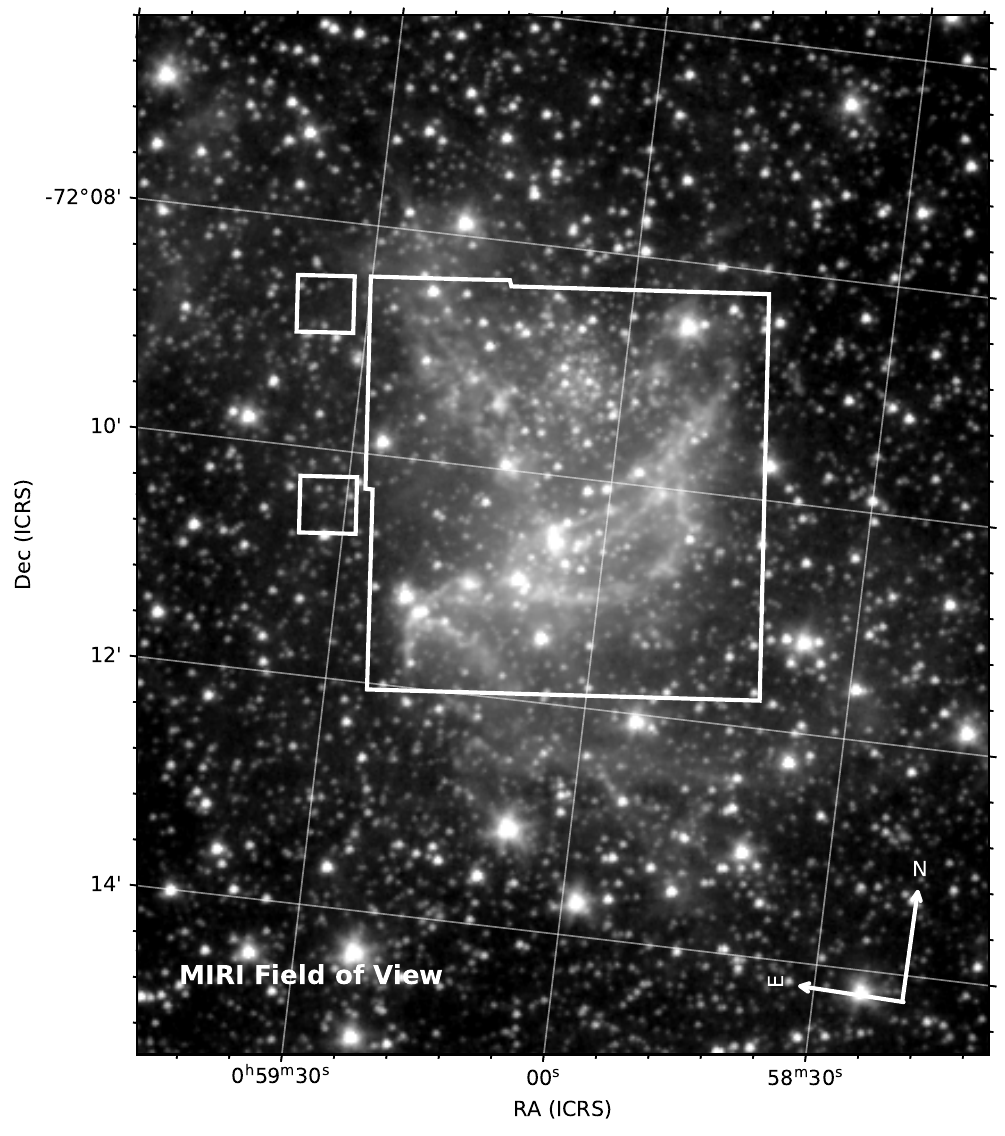}
\end{minipage}
\caption{Location of our {\em JWST} NIRCam (\textbf{left}) and MIRI (\textbf{right}) footprints superimposed on a {\em Spitzer} 3.6 $\mu$m mosaic of NGC 346 \citep{Meixner2006}. North is oriented upward and east to the left. Our NIRCam observation was constructed in two parts. The larger region in white contains three pointings using both the A and B modules of NIRCam. To complete the mosaic, a fourth pointing, also using both the A and B modules (\textbf{shown in yellow}) was obtained using a longer integration time to facilitate simultaneous MIRI imaging off-source.}
\label{fig:footprint}
\end{figure*}

We have imaged the SMC star-forming region NGC 346 in the near- to mid-IR using {\em JWST}'s Near Infrared Camera (NIRCam; \citealt{rieke05, rieke23}) and Mid-Infrared Instrument (MIRI; \citealt{rieke15, Wright2023}). These observations were conducted as part of {\em JWST} GTO program \#1227 (PI:\ M.\ Meixner). The regions of NGC 346 showing active star formation cover $\sim$30 arcmin$^2$ ($\sim$9700 parsec$^2$). With NIRCam, we exposed for a total of 0.93 hours, obtaining data in six filters covering an area indicated in \autoref{fig:footprint}. With the MIRI imager \citep{Bouchet2015, bib:Dicken2024_miri}, we exposed for 3.25 hours in five wavelength bands, covering a smaller region of $\sim$12 arcmin$^2$ ($\sim$4000 parsec$^2$) which overlaps with our NIRCam imaging and encompasses the majority of NGC 346, including the areas where star formation is most active. The pointings of our observations were guided by previous IR imaging surveys \citep[e.g.,][]{Bolatto2007, Gordon2011, Meixner2013} which established the locations of active star formation. Our multi-instrument approach was designed to reveal the structure and composition of the diffuse emission permeating the star-forming region, as well as allow for photometric analysis of the populations within, particularly the younger populations of YSOs. These primary observations of NGC 346 were also conducted in parallel with additional imaging in both MIRI and NIRCam of off-source regions.

\subsection{NIRCam Observations}
\label{sec:obs_nircam}
%NIRCam obs date:2023-05-17T17:40:10.944
Our NIRCam observations were obtained on July 16th, 2023 and consist of six tiled mosaics imaged with the filters F115W, F187N, and F200W in the short-wavelength (SW) channel and F277W, F335M, and F444W in the long-wavelength (LW) channel. This selection of filters was chosen based on several factors: for comparison with past works at similar wavelengths revealing specific chemical signatures at particular wavelengths, and to achieve broad coverage over near-IR wavelengths for photometric analyses. {\em JWST}'s F115W and F200W filters are roughly equivalent to Johnson \emph{J} and \emph{K}s. F187N is sensitive to \paa ~emission.  
{\em JWST}'s F444W is equivalent to the \textit{Spitzer} IRAC $[4.5]$ filter, and F335M partially overlaps with the broader IRAC $[3.6]$ filter. These choices allow for comparisons with previous studies of the IR structure and populations of similar star forming regions, such as the Surveying the Agents of a Galaxy's Evolution (SAGE) survey of the Large Magellanic Cloud (LMC) and SMC \citep{bib:Blum2006, Meixner2006, Simon2007, Whitney2008, Gordon2011, bib:sewilo2013}, and the DUST in Nearby Galaxies with Spitzer (DUSTiNGS) survey  \citep{bib:Boyer2015a, bib:McQuinn2017, bib:Goldman2019}.
Additionally, these selections will be advantageous in future work comparing NGC 346 with ongoing {\em JWST} studies of additional low-metallicity star-forming regions such as N79 (Nayak et al., submitted), 30 Doradus \citep{Sabbi2016, DeMarchi2016, bib:fahrion2023b}, Spitzer I in NGC 6822 \citep{bib:lenkic2023, bib:nally2023} and I~Zw~18 \citep{bib:Hirschauer2024}.

Our observations make use of both the A and B NIRCam modules and the {\sc full} subarray in order to access the widest field of view per pointing. For each band, we constructed our mosaic using four tiles to create a rectangular field of view oriented roughly vertically with respect to declination. (See \autoref{fig:footprint}.) We overlap our tiles such that no gaps are present over the field of view, with row and column overlaps of 5.0\% and 60.0\%, respectively, except for those between the detector subarrays in the SW bands. We center this field of view at RA = 00:59:04.9451, Dec = -72:10:9.15, and cover an area of $\sim$30 arcmin$^2$. Using the {\em JWST} Astronomer Proposal Tool, we employed a dither pattern of {\sc none}, but with {\sc standard} four sub-pixel dither positions and the {\sc bright2} readout pattern for each tile.

Three of the four tiles in each filter were observed using two groups per integration for a total of 171.8 seconds per filter, with the fourth tile observed with seven groups per integration and a total of 601.3 seconds per filter, which enabled the inclusion of a coordinated imaging parallel utilizing MIRI. A summary of exposure parameters can be found in \autoref{tab:nircam_obs_summary}. The fourth tile, in the northwest of the mosaic, was taken simultaneously with MIRI imaging which observed a field off-target from NGC 346. The integration time allocated to this fourth NIRCam tile was therefore determined based on the minimum recommended observing pattern strategy for the simultaneous MIRI observation. Therefore, the number of groups per integration for this single tile was increased to seven.

%\vspace{1 cm}
\subsection{MIRI Observations}
%Our MIRI observations
\label{sec:obs_miri}

We imaged NGC 346 using the MIRI instrument on October 10th, 2023 across five wavelength bands. Our observations consist of six (2$\times$3) MIRI tiles to form a mosaic centered at RA = 00:59:04.9451, Dec = -72:10:09.15, covering the central region of NGC 346 and encompassing the most active sites of star formation with diffuse dust and gas. We adopted a position angle constraint of 5 degrees and included a 10\% overlap of the tiles to ensure gapless coverage. The resulting mosaic covers a total footprint of $\sim$12 arcmin$^2$, and falls within the footprint of our NIRCam observations, as shown in \autoref{fig:footprint}, except for the two small regions imaged through the Lyot coronagraph portion of the MIRI sensor. In total, $\sim$40\% of the region imaged by NIRCam has also been imaged by MIRI.

For these observations, we employed a four-point {\sc cycling} dither pattern and obtained images in five wide-band filters: F770W, F1000W, F1130W, F1500W and F2100W. We list our observing parameters in \autoref{tab:miri_obs_summary}. Our filter selection is informed by \citet{Jones2017a} who predicted the mid-IR colors of various stellar populations in nearby galaxies. This approach is the same followed by \cite{bib:nally2023} in their {\em JWST} study of nearby local-group galaxy NGC 6822. With this filter selection, we are sensitive to the youngest, dustiest objects in NGC 346. Comprehensive multi-wavelength studies of galactic star-forming regions have revealed that a significant portion of YSOs are not detected at near-IR wavelengths \citep{bib:habel21,bib:furlan2016}, thus with MIRI we may access the youngest, more embedded population of YSOs inaccessible to NIRCam. Two of our filters, F770W and F1130W, are also sensitive to dusty PAH emission likely prevalent in this star-forming region. The F1000W filter is similarly sensitive to the 10~\micron{} silicate feature, which is known to be seen in absorption in the SEDs of YSOs enshrouded in the most dense dusty envelopes. 

In a method similar to that implemented for our NIRCam observations, a NIRCam coordinated parallel field was observed at the same time as one of the six MIRI tiles for an off-target background region. However, in the case of MIRI, this tile has identical exposure parameters as the other five.

\newpage

\section{Data Processing \& Images}
\label{sec:processing}

\subsection{NIRCam Data Processing}
\label{sec:processing_nircam}
We processed our uncalibrated NIRCam images using version 1.10.2 of the {\em JWST} pipeline with Calibration Reference Data System (CRDS) version 11.16.20 and CRDS context 1088 (\texttt{jwst\_1088.pmap}) for both Stage-1 and Stage-2 processing.
We introduced an additional correction on the calibrated level-2 images for 1/f noise using the tool {\sc image1overf.py} from \citet{bib:1fcor}.

Because we intended to perform our photometric extraction on each individual dither and not on the combined mosaic, we required each frame to have correct astrometric alignment. To achieve this, we made use of the {\em JWST}/{\em Hubble} Alignment Tool (\texttt{JHAT}; \citealt{bib:rest2023}) as we found direct alignment to {\em Gaia}~DR3 via the \texttt{tweakreg} option of the pipeline insufficient. We used a bootstrapping approach to align our multiple filters. First, we aligned the frames from our F277W filter using a catalog of {\em HST}-detected sources in NGC 346 from \cite{DeMarchi2011} which have been aligned to {\em Gaia}~DR3. This catalog spans the central star-forming region of NGC 346 and contains a similar density of resolved sources as that seen by {\em JWST}. Next, we assembled a complete mosaic image in the F277W filter using Stage-3 of the {\em JWST} pipeline, turning the {\tt tweakreg} step off. This final pipeline step also performs source detection and generates a source catalog. We then fed this resulting F277W pipeline catalog back into \texttt{JHAT}, aligning the individual frames of the remaining filters. This bootstrapping approach was necessary as several individual frames from the SW channel situated farthest from the center of NGC 346 contained no sources in the {\em HST} catalog from \cite{DeMarchi2011}. 
However, because the LW channel of NIRCam has larger subarrays, each frame contained sufficient sources from the {\em HST} catalog, allowing us to align all frames of the mosaic.
Finally, as with our aligned F277W frames, we constructed mosaics for the remaining filters with Stage-3 (\texttt{Image3Pipeline}) of the {\em JWST} pipeline with the {\tt tweakreg} step turned off.

\subsection{MIRI Data Processing}
\label{sec:processing_miri}

We generated calibrated MIRI mosaics using {\em JWST} pipeline version 1.9.5 with CRDS version 11.16.21 and context 1077 (\texttt{JWST\_1077.pmap}). We processed each uncalibrated MIRI frame through \texttt{Dectector1Pipeline} and \texttt{Image2Pipeline} with default parameters. The resulting images were then aligned to {\em Gaia}~DR3 using the \texttt{tweakreg} step in the pipeline, before combining them into mosaics using the Stage-3 (\texttt{Image3Pipeline}) step. We accounted for the background in the mosaics by determining and subtracting the median background level from a region free of diffuse emission from NGC~346 to the east. Examples of the resulting mosaics are shown in \ref{fig:compare_nircam_miri} and \autoref{fig:three_color}.

For the F1500W and F2100W filters, only minimal structure around the brightest sources remained in the median background. Even though the residual contamination was low, it could still manifest itself in background subtracted images as faint shadows dispersed on the image. To remove the residuals we manually masked their location and applied an approximate `filling' of the gaps using the \texttt{ndimage.distance\_transform\_edt} module in SciPy. Once backgrounds were created for all filters, we subtracted the instrumental background images from each dither in the mosaic tiles. 
Overall we found that our background subtraction methods performed well in reducing the background level across all mosaic tiles in all filters. We then constructed mosaics from the background subtracted images using \texttt{Image3Pipeline}.

\subsection{Images} 
\begin{figure*}
\centering
\includegraphics[trim=0cm 0cm 0cm 0cm, clip=true,width=\textwidth]{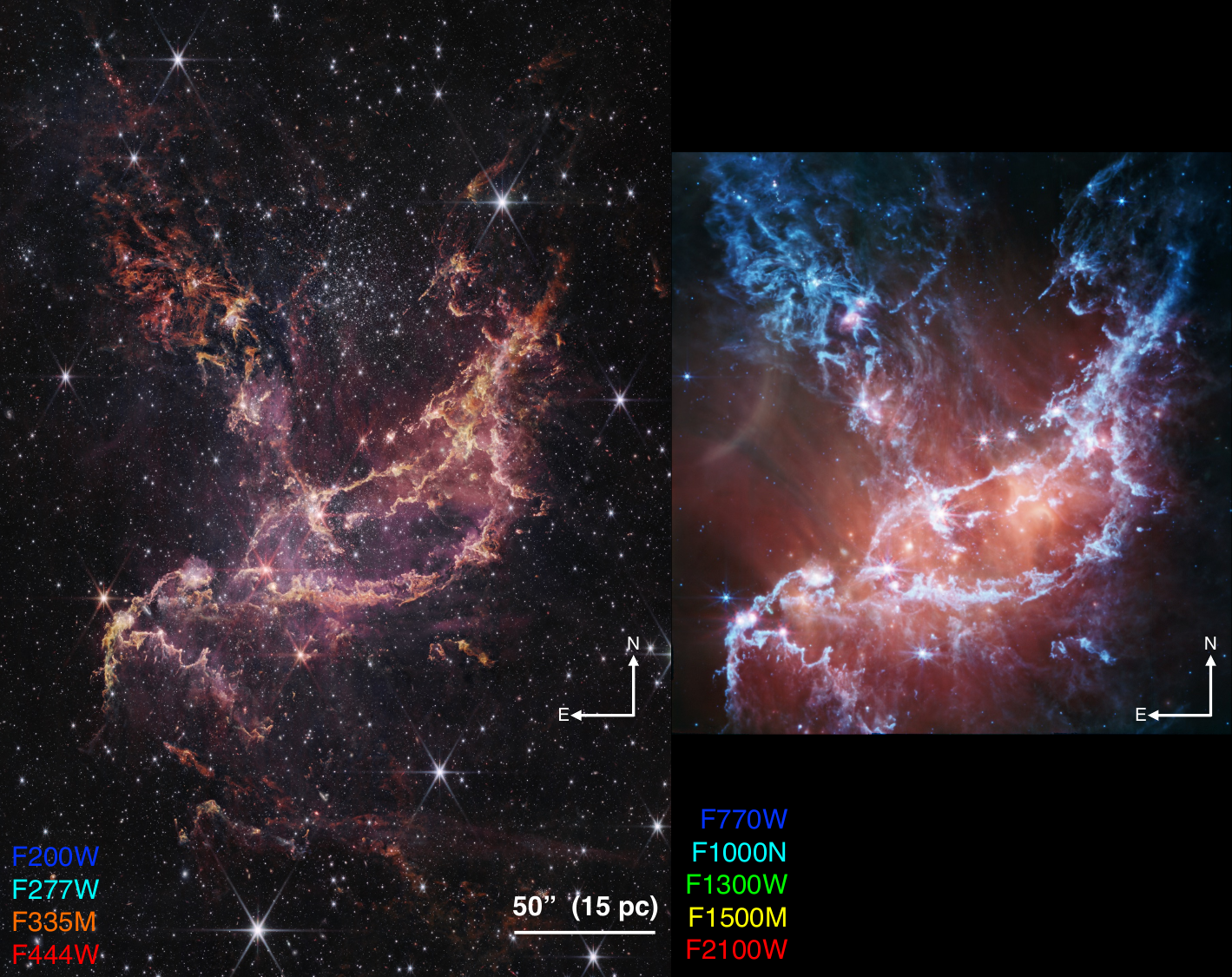}
%\vspace{3cm}
 \caption{Multi-color images of NGC 346. Credit: NASA, ESA, CSA, Olivia C. Jones (UK ATC), Guido De Marchi (ESTEC), Margaret Meixner (USRA); NIRCam processing: Alyssa Pagan (STScI), Nolan Habel (USRA), Laura Lenkić (USRA), Laurie E. U. Chu (NASA Ames); MIRI processing: Alyssa Pagan (STScI), Patrick Kavanagh (Maynooth University).}
 \vspace{0.5 cm} %keep this one
 \label{fig:compare_nircam_miri}
 \end{figure*}

Imaging across eleven different wavelength bands reveals the diverse stellar populations and diffuse structure present in NGC 346. NIRCam imaging (\autoref{fig:compare_nircam_miri}; \textbf{left}) reveals a dense population of stars across the field, including the intermediate-age cluster BS90, located in the northwest region of the field \citep{bib:bica1995}. Complex filamentary structure traces out the main ``arc" of NGC 346 where emission by warm dust and PAHs (as seen in the F335M and F770W bands) dominate. MIRI imaging (\autoref{fig:compare_nircam_miri}; \textbf{right}) further reveals the complex diffuse emission in the region. At mid-IR wavelengths, far fewer individual stars can be seen as emission at these wavelengths in evolved sources begins falling rapidly. However, MIRI reveals several prominent bright red sources which emit more strongly in the mid-IR than near-IR. These sources are likely less evolved and are coincident with areas of dusty diffuse nebulosity.

\begin{figure}
\centering
%\vspace{0.5 cm}
\includegraphics[width=\columnwidth]{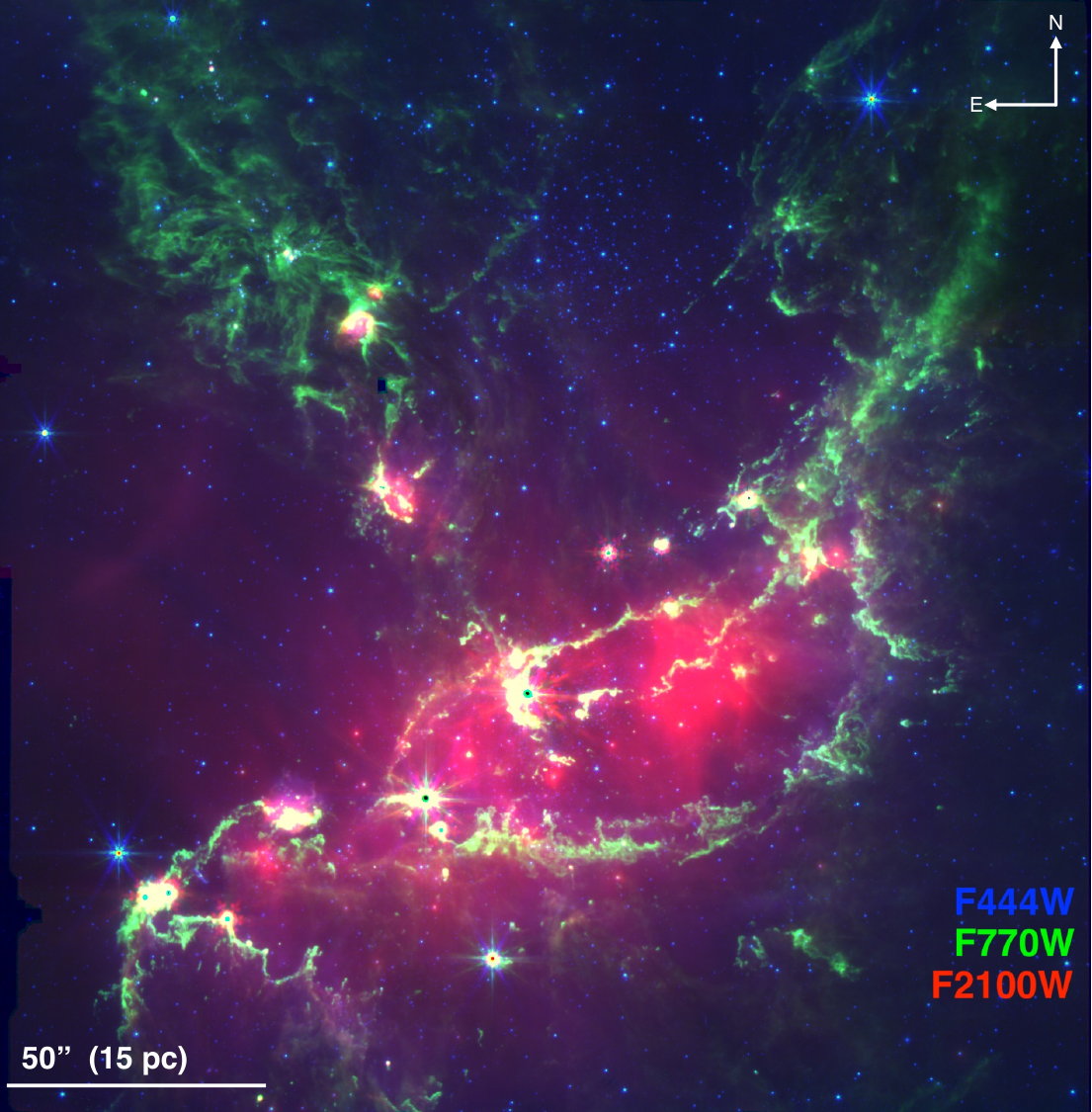}
\caption{A three-color image of NGC 346 combining observations from NIRCam and MIRI. Imaging taken with the F444W filter is colored blue and highlights the more evolved stellar population in this region including the cluster BS90 in the upper portion. F770W, colored green, traces dusty PAH emission, prominent in the northeast and western regions. F2100W reveals warm, ionized gas and hot dust in the central region.} 
\label{fig:three_color}
\end{figure}

Combining wavelengths across both instruments clearly reveals the stark contrast between the various components of NGC 346. In \autoref{fig:three_color} we combine imaging from the F444W (blue), F770W (green) and F2100W (red) bands. At 4.4~$\mu$m, the stars in the field remain visible, with the more evolved sources such as those in the BS90 cluster appearing blue. Emission at 7.7~$\mu$m traces PAHs which lie at the interface between ionized and molecular gas. Warm, ionized gas and hot dust is traced at 21~$\mu$m, notably surrounding many areas of the central region associated with active star formation. At these locations, the less evolved stars emit more at longer wavelengths than their older, bluer counterparts, appearing as white or red.

\label{sec:images}

\newpage

\section{Source Extraction and Photometry}
\label{sec:photometry}
For extracting photometry from our NIRCam and MIRI observations, we employ {\sc starbugii} \citep{Nally_Starbug2_2023}, a Python-based photometric analysis tool built for {\em JWST} imaging data with capabilities specifically tailored to source extraction in fields containing complex diffuse emission, a historically-difficult challenge for studies of star-forming regions. This tool is capable of both aperture and PSF photometry of NIRCam and MIRI imaging data of any given filter band, calculating zero-point corrections, and merging photometric catalogs from multiple bands into a single source matched catalog. Here we discuss our methods for source detection and constructing a photometric catalog of NIRCam PSF photometry and MIRI aperture photometry.

\subsection{Source Detection and Aperture Photometry}
\label{sec:source_detection_aperture_phot}
We conduct source detection on our NIRCam and MIRI data similarly. For all filters in both instruments, we use {\em Gaia}-aligned Stage-2 images and the command \texttt{starbug2 $--$detect}. This routine finds sources with a 5$\sigma$ detection above the locally-estimated sky level and applies geometric \textit{sharpness} and \textit{roundness} limits, helping to remove cosmic rays and eliminate resolved background galaxies.

Our observation scheme when using NIRCam resulted in a minimum coverage of four exposures at any given location within our footprint, save for the outer edges or the narrow gaps between the SW subarrays. Thus we demanded that sources be detected in at least three dithers for inclusion into our catalog. Similarly, in MIRI, our observations contained four exposures at a given pointing, and we again required a detection in at least three dithers for inclusion. This strict matching criteria was intended to combat the occurrence of sporadic sources such as cosmic rays or detector artifacts and achieve confident matching of true sources between exposures. The {\sc starbugii} parameters used in our source detection and source matching are shown in \autoref{tab:sb_params}.

For both our NIRCam and MIRI images, we conduct aperture photometry for the sources found in our detection. We adopted a fixed aperture radius of 1.5 pixels with a sky annulus of inner and outer radii of 3.0 and 4.5 pixels respectively. 
Aperture corrections are calculated and applied from given CRDS values in \texttt{jwst\_nircam\_apcorr\_0004.fits} for NIRCam and \texttt{jwst\_miri\_apcorr\_0005.fits} for MIRI. This aperture photometry is similar to that used in \citet{bib:jones2023}, which also conducted aperture photometry for this same set of NIRCam observations using {\sc starbugii}. Here, however, we have improved upon their efforts by re-processing the original data with newly-delivered image calibrations and improved its astrometric alignment with the help of \texttt{JHAT} \citep{bib:rest2023}.

\subsection{PSF Photometry} 
\label{sec:psf_photometry}
In addition to the aperture photometry described in \autoref{sec:source_detection_aperture_phot}, we conducted PSF-fitted photometry on our NIRCam observations. We employ \texttt{ starbugii \-\-background} to model the diffuse, nebulous emission present in NGC 346. This routine, described in further detail in \cite{bib:nally2023}, masks detected sources to create a representation of the nebulous emission devoid of stars which is then subtracted from the original exposures, leaving a nebulosity-free image upon which PSFs may be fitted.
Using \texttt{ starbugii \-\-psf}, we generated a $5''$ radius PSF from {\sc webbpsf}~\citep{webbpsf2014} version 1.1.1 for each detector subarray in NIRCam. Once a diffuse background is generated and subtracted from every frame, we use the source catalog from our aperture photometry to provide initial coordinate guesses for our PSF fitting. The fitting routine of {\sc starbugii} then fits a new centroid position and flux, if necessary force-fitting on the initial guess if no new centroid can be found within a $0\farcs1$ radius.

We defer similar PSF fitting for the MIRI component of our observations because of limitations of the MIRI PSFs simulated by {\sc webbpsf} at the time of processing. These simulated PSFs did not include the cruciform structure known to contribute to up to $\sim$26\% of the flux in the F770W band \citep{bib:Gaspar2021}. Experimentation in MIRI PSF photometry yielded poor fitting and systematic errors in the form of lost flux. Similarly, we note difficulties in fitting PSFs for the F200W filter. PSF photometry in this band showed unreliable photometry results, particularly in regions where our mosaic tiles overlapped. In CMDs constructed using F200W and another band, this manifested as multiple population sequences systematically offset from one another in the F200W photometry indicating the loss of some flux. Thus in this work we use aperture photometry for the F200W filter at the cost of some loss in astrometric precision. 

We note this combined aperture-PSF catalog for NIRCam shows improvements in the quality of source matching upon the aperture-only catalog of \citet{bib:jones2023}. 
Though \cite{bib:jones2023} reports a total of more sources detected in at least one NIRCam band, we assess that this difference is caused by a greater degree of mismatching attributable to looser matching constraints and lesser astrometric precision. Evidence of this can be observed in the CCDs and CMDS of \cite{bib:jones2023} which show some horizontal and diagonal scatter extending from regions of high source density, an effect characteristic of erroneous matching of detections between dithers and of sources between bands. This is reinforced by making a similar comparison of our F115W photometry with that of the optical \hst ~catalog from \cite{Sabbi2007} and \cite{DeMarchi2011} as was made in \cite{bib:jones2023}. While \cite{bib:jones2023} report 24,367 sources in common, we find a greater number of common sources (24,939) using the same matching radius of $0\farcs3$. This represents $\sim$85\% of the sources from \cite{Sabbi2007} and \cite{DeMarchi2011} which is approximately the overlap in the field of view of their \hst ~observations with our NIRCam imaging. Finally, the CCDs and CMDs we show in \autoref{sec:results} are free from the scattering artifacts in the source catalog of \cite{bib:jones2023}.

In MIRI, the reduced crowding and lower overall source count make the improvement in positional fitting and precision in flux measurement offered by PSF photometry less important than for NIRCam. However, future work will implement updated model PSFs for both NIRCam and MIRI when available.

\subsection{Photometric Corrections} 
\label{sec:corrections_photometry}
To determine a zero-point correction to our NIRCam PSF photometry we first create a clean subset of each filter's aperture photometry catalog by eliminating the brightest and faintest sources, those approximately one magnitude away from our detection limits on either end. This removes most sources with low S/N and those that are saturated. We additionally remove sources with photometric errors larger than 0.1 magnitudes as determined by {\sc starbugii} and any sources with data flags indicating poor quality. We then employ \texttt{starbug2 $--$calc-instr-zp} to apply our zero-point corrections. This routine matches the cleaned aperture photometry catalogs with their given PSF catalogs and calculates the median difference to find each filter's instrumental zero-point. These corrections are then applied to the entire PSF catalog for each filter. Finally, we convert both our NIRCam and MIRI catalogs from AB to Vega magnitudes using the CRDS files {\sc jwst\_nircam\_abvegaoffset\_0001.asdf} and {\sc jwst\_miri\_abvegaoffset\_0001.asdf}.
Because NGC 346, located in the SMC, is not affected by significant Galactic reddening especially at mid-IR wavelengths, we do not apply an additional reddening correction to our catalogs. 

\subsection{Source Catalog and Completeness}
\label{sec:catalog}
After applying the corrections mentioned in the previous section to each individual filter catalog, we combined them into a single source-matched catalog. To prepare each filter catalog, we eliminated sources with a S/N of 10 or lower, which reduced mismatching of sources between filters, particularly in regions dense with detected sources, many with low-quality detections. We note that this choice has implications for the photometric completeness of our catalog at the faint limit of our detections. We emphasize that in this study, our goal is to characterize the physical properties of the sources with small photometric uncertainties particularly the young sources which have been detected across multiple filters, rather than  characterizing the  very faintest sources in the field which are less likely to be detected and matched across multiple filters. 

\begin{figure}
\centering
\includegraphics[width=\columnwidth]{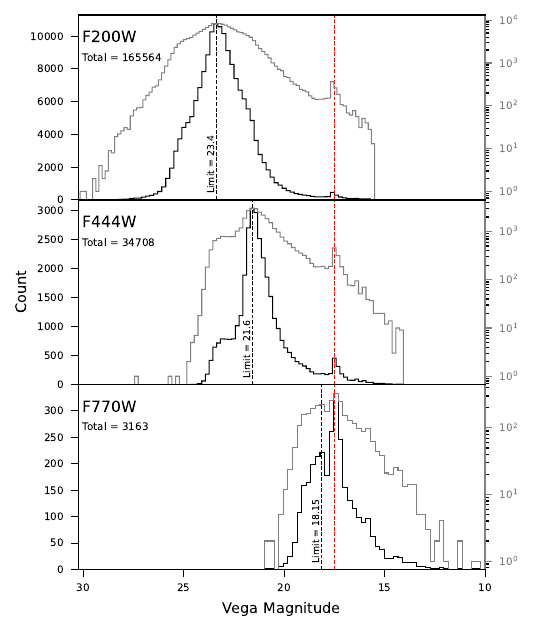}
\caption{Luminosity functions of sources detected in the F200W, F444W and F770W filters. Linear and logarithmically scaled distributions are outlined in black and gray respectively. The black dashed line marks the inferred completeness limit. The red dashed line marks the location (17.5 magnitudes) of the RC for these filters.} 
 \label{fig:luminosity_functions}
 \end{figure}

We employ luminosity functions to assess the completeness of our catalog. In \autoref{fig:luminosity_functions} we show examples of luminosity functions for bands in the SW NIRCam channel (F200W), the LW NIRCam channel (F444W) and in MIRI imaging (F770W). For each wavelnth band, we have selected optimized bin widths using Knuth's Rule \citep{bib:knuth2006}. For each filter, we have estimated a completeness limit by locating the turnover of our luminosity functions toward the faint limit of our histograms to within 0.05 magnitudes, illustrated by the black dashed line in \autoref{fig:luminosity_functions}. In these luminosity functions, we detect for wavelengths between $1.2-10$\micron, evidence of the RC, which appears as a secondary peak at the bright end of the distributions centered approximately at $17.5$ magnitudes across these wavelengths. (Note that this peak occurs at the fainter magnitude of $\sim 18.5$ magnitudes for the F115W in line with the expectation that the RC appear fainter towards bluer near-IR wavelengths.) In our F770W band photometry, we observe that the RC composes a significant portion of the sources detected, resulting in a maximum at $17.5$ magnitudes adjacent to a fainter peak where the completeness fall-off begins. This two-peak distribution is consistent with CMD analysis described in \autoref{sec:results}. We summarize the sensitivity of all filters in \autoref{tab:sensitivity}. The uneven integration times across our NIRCam mosaic result in slower fall-off in the luminosity function on the fainter side of the inferred completeness limit. Because one pointing pointing (shown in yellow in \autoref{fig:footprint}) reached a deeper integration, a greater number of fainter sources are detected in these regions. This feature is most prominent in the F444W luminosity function as a shelf-like feature with a slight secondary peak at $\sim24$ magnitudes.

 \begin{table}
    \centering
        \caption{Source counts and sensitivity as a function of filter. Our completeness is estimated to within 0.05 magnitudes based on the turnover of our luminosity functions. We report the brightest limits found in each filter, above which our images become saturated. Values are reported in Vega magnitudes.}
    \begin{tabular}{lrcc}
        \hline
        \hline
         Filter & Source & Completeness & Saturation  \\
                & Count  &   [VegaMag] & [VegaMag] \\
         \hline
         \hline
        F115W &  131443  & 23.75 & 13.56 \\
        F187N &  13557   & 19.95 & 12.30 \\
        F200W &  165564  & 23.40 & 15.50 \\
        F277W &  87638   & 22.85 & 13.31 \\
        F335M &  52966   & 21.75 & 13.95 \\
        F444W &  34708   & 21.60 & 14.07 \\
        F770W &  3163    & 18.15 & 10.22 \\
        F1000W & 1914    & 17.40 & 9.05 \\
        F1130W & 1170    & 15.35 & 8.52 \\
        F1500W & 605     & 14.70 & 7.43 \\
        F2100W & 261     & 12.15 & 6.70 \\

         \hline
         \hline
    \end{tabular}
    \label{tab:sensitivity}
\end{table}

We used \texttt{starbug2-match $--$band} in order to merge our individual filter catalogs. This routine first treats the two instruments separately, creating a band-matched catalog for NIRCam and another for MIRI. For both instruments, the routine begins with the shortest wavelength bands as the PSFs with smaller full width at half-maximum (FWHM) are assumed to have greater astrometric precision. The remaining filters are sequentially matched using a nearest-neighbor method. When a source cannot be matched to another within the matching radius, it is appended to the catalog as a new source. We adjust the matching radius for each filter to account for increasing astrometric uncertainty at longer wavelengths. 
For NIRCam's SW filters we adopt a matching threshold of $0\farcs06$, and $0\farcs1$ for its LW filters. For MIRI, we adopt $0\farcs15$ for the F770W, F1000W and F1130W filters; $0\farcs2$ for F1500W; and $0\farcs25$ for F2100W.

In the final step, we combine the NIRCam and MIRI catalogs. The catalogs are joined by comparing detections made in the F444W band in NIRCam and the F770W band in MIRI using a matching radius of $0\farcs3$. Because our field contains many populations which may be brighter or fainter at the extrema of our wavelength range or not even detected, using the nearest adjacent filters ensures that detections paired together are more likely to be attributable to the same object. This is particularly necessary for the most embedded protostellar sources which are expected to appear bright at mid-IR wavelengths but not luminous enough in the near-IR to be detectable. To further combat erroneously matching a source detected in the mid-IR with an unrelated object in our more-crowded near-IR observations, we also require that a source detected in MIRI have a F444W counterpart, that is, if no F444W counterpart can be found for F770W, the algorithm will not search for a match in any shorter-wavelength filters. It will instead append the source to the catalog as containing MIRI photometry only and no photometry for NIRCam.
The final coordinates of each source in the combined catalog are adopted from the shortest wavelength filter for which a detection exists which is assumed to have the greatest astrometric accuracy.

%------------------------------------------------------------------------------------------------

\section{Results}
\label{sec:results}

\subsection{NIRCam Color-Magnitude and Color-Color Diagrams}
\label{sec:nircam_cmds_ccds}

\begin{figure*}
\centering
\includegraphics[scale=.8]{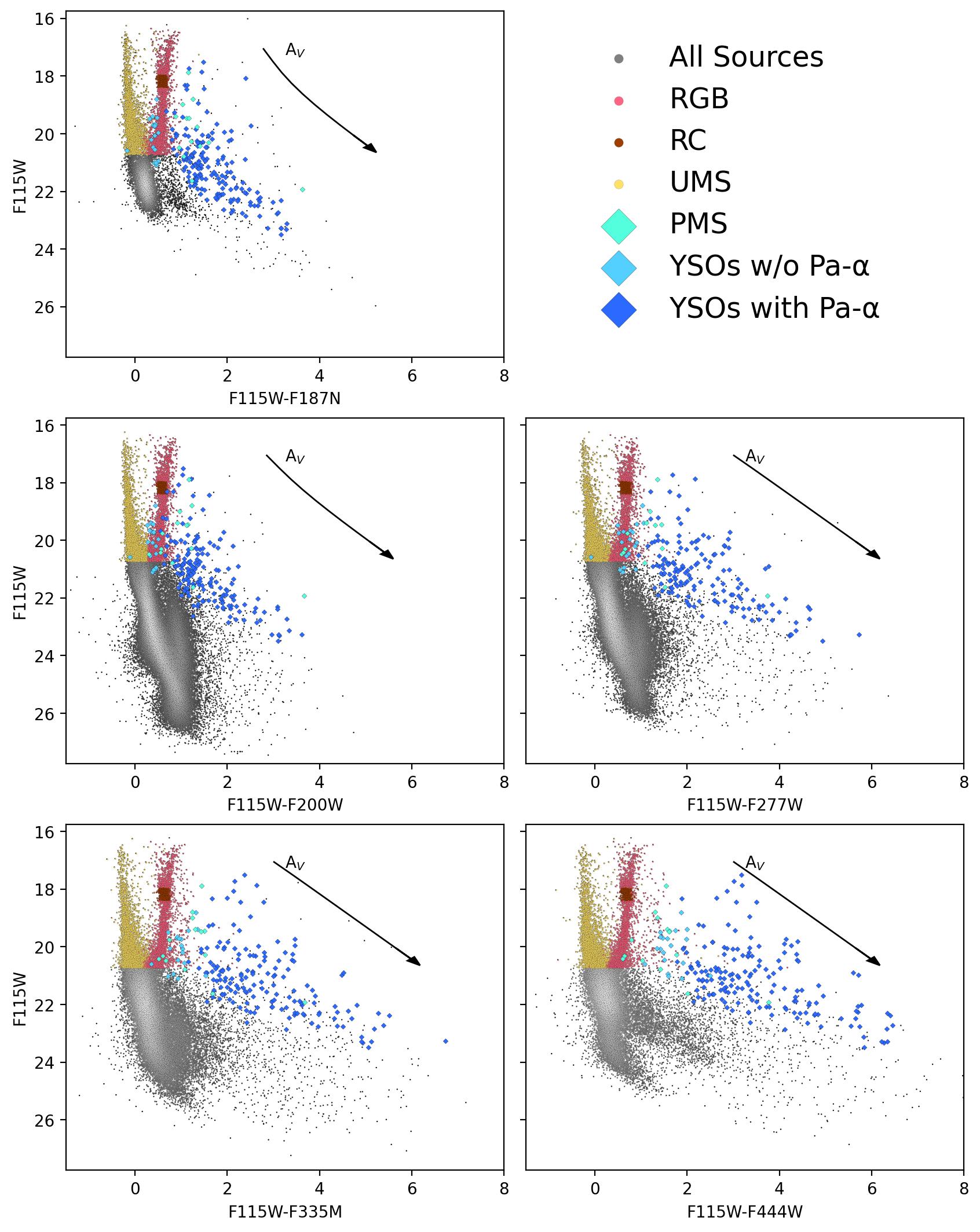}
%\vspace{0cm}
 \caption{Five CCDs created from combinations of our six NIRCam filters.  
 We clearly observe a main sequence with an RGB and a tightly clustered RC with a magnitude of $\sim$18.25 magnitudes in F115W. A significant number of sources display an IR excess, appearing to the right of the MS and RGB. NIRCam-identified YSO and PMS sources are marked as shown in the legend. The black curve traces extinction in A$_V$ from of $0$ to $23.2$ magnitudes.} 
 \label{fig:cmd_nir}
 \end{figure*}

To characterize the properties of the stars in our field, and to identify pre-stellar populations in NGC 346, we employ CMDs and CCDs. \textcolor{red}{\autoref{fig:cmd_nir}} shows several CMDs where the F115W NIRCam photometry is compared with increasingly redder colors. In the F115W band, the entire CMD extends from $\simeq 28$ magnitudes at the faintest to $\simeq 17.5$ at the brightest above which sources begin to saturate. 
One can discern a clear main sequence (MS; grayscale dots), fainter than F115W $\simeq 21$, which at brighter magnitudes bifurcates into an upper main sequence (UMS; yellow dots), and a red giant branch (RGB; dark red dots). Within the RGB itself, one can clearly detect the red clump feature (RC; dark red dots) as an overdensity of sources clustered together at F115W $\simeq 18.2$ (see reviews by \citealt{bib:Girardi2016} and \citealt{bib:Onozato2019}.)
See \autoref{tab:nircam_source_table} for the color and magnitude limits used to separate the different populations.
For each CMD, we trace the reddening curves across an arbitrary range of values of A$_V$ spanning $\sim23$ magnitudes in order to visualize the impact of extinction. We calculate these values of A$_V$ using the Python synthetic photometry package \texttt{synphot} using the SMC Bar Average Extinction Curve described in \citet{bib:gordon2003}. The curvature results from the variation of effective wavelength as reddening increases, and is particularly pronounced when one combines the wide-band F115W filter with the narrow-band F187N. The lack of noticeable elongation of the RC feature in particular indicates that in the NGC 346 field, the effect of differential reddening is marginal, for comparison less than that found in the LMC region 30 Doradus which showed a color spread of $>1$ magnitude in the F110W and F160W filters of {\em HST} \citep{DeMarchi2014, Sabbi2016, DeMarchi2016, bib:fahrion2023b}.

Note that \citet{bib:jones2023} detected a more noticeable elongation of this feature in these CMDs created from aperture photometry alone. We attribute this discrepancy to the recent improvements to the {\em JWST} image calibration, as well as the greater capabilities of PSF photometry to accurately measure fluxes in crowded fields.

\begin{table*}[h]
\centering
\begin{threeparttable}
\caption{NIRCam-Identified Populations in NGC 346}
\label{tab:nircam_source_table}
\begin{tabular}{lcr}

\hline
\hline
Population & Color Criteria & Number of Sources   \\
\hline
\hline
UMS   				& $-0.37 <$ F115W--F200W $< 0.27$ and F115W $< 20.71$ 	& 3166   \\
RGB  				& $0.27 <$ F115W--F200W $< 1.02$ and F115W $< 20.71$ 	& 3142   \\
RC   				& $0.45 <$ F115W--F200W $< 0.68$ and $17.94 <$ F115W $< 18.37$ & 546	\\
\hline
YSOs with \paa \tnote{1}  		& F115W--F187N $> 0.5$ and  F200W--F444W $> 0.97$		& 163 \\
YSOs without \paa \tnote{1} 				& FF115W--F187N $< 0.5$ and  F200W--F444W $> 0.97$	& 19 \\
PMS \tnote{1}   				& F115W--F187N $> 0.5$ and  F200W--F444W $< 0.97$		& 14 \\
\hline
\hline
\bf{All Sources}  			& - 							& \bf{203891}   \\
\hline
\hline 
\end{tabular}
\begin{tablenotes}
\item[1] YSO and PMS candidates must also be redder by at least four times the uncertainty of their color than the contour shown in \autoref{fig:ccd_nir} which bounds evolved populations as determined by theoretical isochrones.
\end{tablenotes}
\end{threeparttable}
\end{table*}

A significant population of fainter and redder sources is present, separated from the lower portion of the MS. The extent of this region is increasingly apparent when examining a wider color baseline, such as in the F115W {\em vs.}\ F115W--F444W CMD. 
While in principle a population of the UMS could occupy this color space given extremely high reddening values, the lack of observed reddening as seen in our RC excludes this explanation, indicating that these sources are likely young in nature.

\begin{figure*}
\vspace{-5.0cm}
\centering
\includegraphics[width=\textwidth]{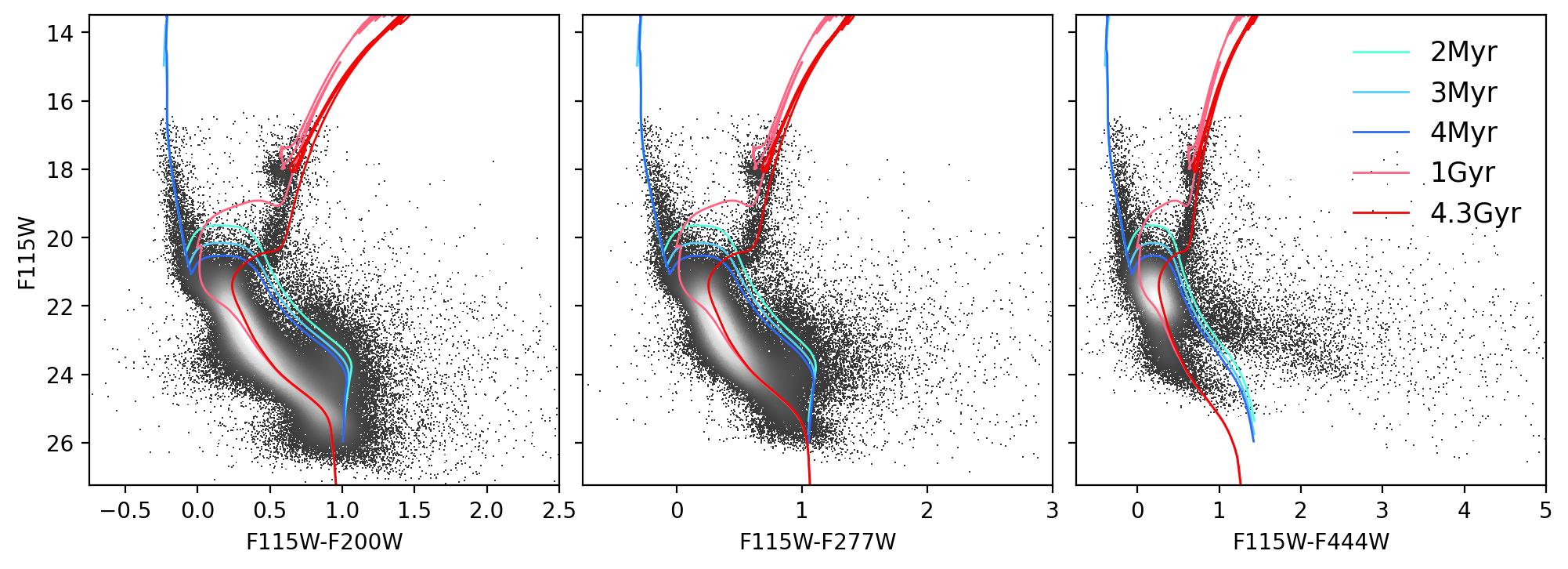}
%\vspace{3cm}
 \caption{Three CMDs showing combinations of our wideband NIRCam filters (also shown in \autoref{fig:cmd_nir}) and isochrones of select ages. We use MIST isochrones with metallicity $[Fe/H]=-1.0$ \citep{Peimbert2000}. Isochrones for 2Myr, 3Myr, 4Myr, 1Gyr and 4.3Gyr are shown in cyan, light blue, dark blue, pink and red respectively. The youngest three isochrones follow the youngest stellar population in NGC 346 found by \cite{Sabbi2007} with \hst ~($\sim 3\pm1$Myr). The isochrone of ($4.3\pm1$Myr) corresponds to the age of the foreground intermediate-age SMC star cluster BS 90 as determined also by \cite{Sabbi2007}.}
 \label{fig:cmd_iso}
 \end{figure*}

In \autoref{fig:cmd_iso}, we show CMDs created comparing our wide band filters with the F115W filter and compare them with a number of select isochrones. We display isochrones from the MIST stellar models to help illustrate the populations present in these three CMDs \citep{bib:MIST}. Our isochrones are created assuming a metallicity of $[Fe/H]=-1.0$ from \cite{Peimbert2000}
an extinction ($A_{\rm V}$) of 0.12 magnitudes (as the photometry is not itself corrected for extinction and may thus may be slightly affected) and finally a distance modulus of $18.96$ \citep{bib:scowcroft2016}. In an \hst ~imaging study of populations in the SMC using the Advanced Camera for Surveys (ACS), \cite{Sabbi2007} found an age of $\sim 3\pm1$Myr for the youngest stellar population associated with NGC 346 and an age of $4.3\pm0.1$Gyr for the foreground intermediate-age SMC cluster BS 90. In \autoref{fig:cmd_iso}, we plot isochrones for the ages 2Myr, 3Myr, 4Myr, 1Gyr and 4.3Gyr. We find that the 2-4Myr isochrones follow the UMS and extend into a reddened population present at a color of $\sim 1$ magnitude for all three CMDs. The older 4.3 Gyr isochrone traces the MS and roughly follows the RGB. We note a small color offset from this isochrone to the center of our RGB and RC distribution of approximately $\sim 0.075$ magnitudes in the F115-F200W {\em vs.} F115W CMD and closer agreement in the F115W-F277W {\em vs.} F115W and F115W-F444W {\em vs.} F115W CMDS. We suggest this offset may be a consequence of inaccuracies in the zero-points and Vega magnitude offsets used in the photometric calibrations of the MIST isochrones which are only current to pre-launch values. Alternatively, it may point to a younger age in this population of stars, down to an age of 1Gyr, the isochrone for which falls in line with the center of our RGB distribution and extends to the center of the RC.

To better characterize the properties and investigate the nature of these sources, we follow a similar approach as \citet{bib:jones2023}, who previously identified objects with near-IR or \paa\ excess from their broad- and narrow-band colors. We note that our goal is not that of identifying all possible objects with excess, but rather to build a catalog of reliable high-S/N sources as candidates for future near-IR spectroscopic follow-up. Therefore, we will only consider objects with small photometric uncertainty.

\begin{figure}
\centering
\includegraphics[width=\columnwidth]{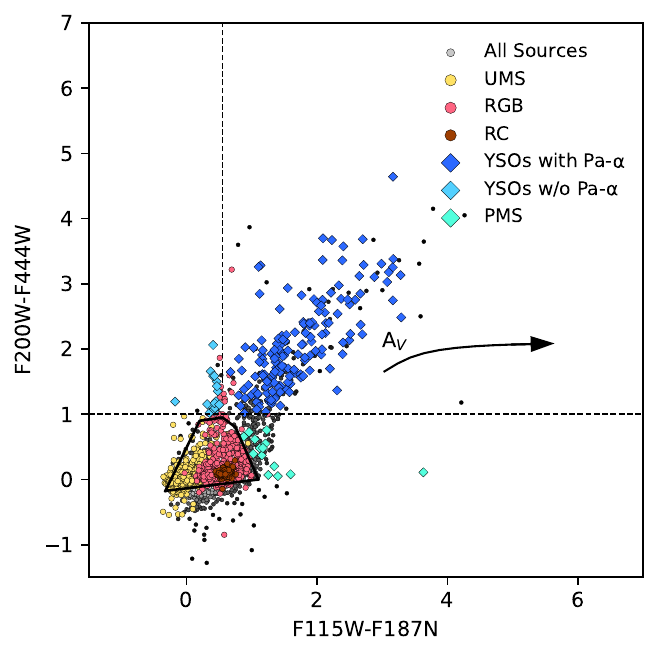}
%\vspace{3cm}
 \caption{A CCD showing stellar populations in NGC 346 identified by NIRCam color-magnitude criteria. We observe a significant population (in dark blue) of YSO candidates showing IR (F200W-F444W) and \paa ~(F115W-F187N) excess. The black contour denotes the region of color-space occupied by evolved objects as determined from theoretical isochones. The black curve traces extinction in A$_V$ from $0$ to $23.2$ magnitudes.}  
 \label{fig:ccd_nir}
 \end{figure}

\autoref{fig:ccd_nir} shows the F200W--F444W {\em vs.}\ F115W--F187N CCD of all sources detected in these four bands. In searching for YSO populations, we consider only those with photometric uncertainty smaller than $0.05$ mag in the F115W and F200W bands. In order to identify objects with significant near-IR or \paa\ excess, we first define the region of the CCD where normal stars are expected to be found, in particular MS, UMS, RGB, and RC stars. This is first done empirically, by looking at the location of the four types of objects defined in \autoref{tab:nircam_source_table}. We then looked at the color range spanned in this diagram by theoretical isochrones for different ages in the four selected bands. In particular, we used as a reference the PARSEC \citep{bib:tang2014} and MIST \citep{bib:dotter2016} theoretical models for metallicity $Z=0.004$ (equivalent to 0.2$Z_\odot$) and for ages of $3, 30, 300$\,Myr, and 3\,Gyr, after having excluded all sources that would appear brighter than F115W = 16 in NGC\,346, since there are no such objects in our photometry. The models occupy a roughly triangular region, consistent with the location of the MS, UMS, RGB, and RC stars. The black region marked in \autoref{fig:ccd_nir} represents a contour containing the area of color-space spanned by the models.

%\begin{figure}
%\centering
%\includegraphics[width=\columnwidth]{CCD_Populations_Revised.pdf}
%%\vspace{3cm}
% \caption{A CCD showing stellar populations in NGC 346 identified by NIRCam color-magnitude criteria. We observe a significant population (in dark blue) of YSO candidates showing IR (F200W-F444W) and \paa ~(F115W-F187N) excess. The black contour denotes the region of color-space occupied by evolved objects as determined from theoretical isochones. The black curve traces extinction in A$_V$ from $0$ to $23.2$ magnitudes.}  
% \label{fig:ccd_nir}
% \end{figure}

All objects with F200W--F444W or F115W--F187N colors larger than those defined by the black region have in principle some excess emission. To select objects with a significant excess, we only consider objects that are redder than the contours of the region by at least four times the photometric uncertainty on the colors.
The selected sources, 196 in total, are marked with diamonds in \autoref{fig:ccd_nir} and are divided in three groups:\ (1) objects with only \paa ~excess (teal) with $>4\,\sigma$ excess in F115W--F187N and F200W--F444W$<1$ (14 sources); (2) objects with Pa$\,\alpha$ and near-IR excess (dark blue), with $> 4\,\sigma$ excess in both F115W-F187N and F200W-F444W and with both colors larger than 1 (163 sources); and (3) objects with only near-IR excess (light blue) with $> 4\,\sigma$ excess in F200W--F444W and F115W--F187N$<1$ (19 sources). Group 1 and 2 sources cannot be generally interpreted as heavily-reddened objects, as the reddening line in \autoref{fig:ccd_nir} is unable to explain the large F200W-F444W color. They present an intrinsic IR excess and are likely PMS stars and YSOs; some YSOs might also be present in Group 3. To investigate this, we turn to the CMDs of \autoref{fig:cmd_nir}, where the same symbols are used. 
The majority of Group 1 sources are fainter than F115W = 21 and are compatible with being PMS stars. (We adopt the label ``PMS" to reference this population in the remainder of this work.) The color distributions of these objects in broad and medium bands place them relatively close to the MS and redder than the RGB. 
These are the positions in the CMD where respectively older and younger PMS stars are expected (see, e.g., \citealt{DeMarchi2010, DeMarchi2011, bib:demarchi2013a, bib:demarchi2013b, bib:demarchi2017}).

Groups 2 and 3 (YSOs with \paa ~and YSOs without \paa ~hereafter) show IR excess in nearly all CMDs. YSOs without \paa ~appear in general less reddened than the larger population of YSOs with \paa, indicating that they may be at a later stage of evolution having a thinner protostellar envelope and experiencing a period of quiescence between periods of active \paa-generating accretion. Accordingly, they show colors closest to 0 in the F115W--F187N CMD of any of the three young populations. YSOs with \paa ~in contrast show significantly-redder colors in every CMD, extending out to 6 magnitudes in F115W--F200W and also extending to fainter magnitudes; approximately 2 below the median brightness of PMS sources and YSOs without \paa ~in each CMD. We note that the CMDs constructed from the bands of longer wavelength than F187N all show a significant population of red sources fainter than those identified by our criteria as some form of YSO. This population very likely contains additional true YSOs which were too faint to be detected in our narrow F187N band, thus not meeting our strict color criteria. Future works taking a more \paa-agnostic approach may characterize this population further and determine their nature.  

\subsection{Reddened MIRI Sources}
\label{sec:miri_red}

Previous studies of nearby extra-galactic star-forming regions have made use of \textit{Spitzer} imaging with IRAC and MIPS to detect YSOs by identifying the location in color-magnitude space in which they lie. Such regions studied include the SMC, the LMC, and NGC 6822, a nearby barred, irregular galaxy in the local group (\citealt{bib:sewilo2013}; \citealt{Whitney2008}; \citealt{bib:Jones2019}, respectively).
Here we follow a similar approach, making use of all of our MIRI bands in identifying regions in color space containing reddened, YSO-candidate sources. We constructed ten CMDs of filter combinations shown in \autoref{tab:MIRI_color_cuts}.  Combinations spanning the shortest mid-IR bands (e.g., F770W--F1000W and F770W--F1130W) are sensitive to sources which are redder, but may not be seen at the longest wavelength, such as an intermediate or later-stage YSO with a depleted envelope. In contrast, combinations spanning the longest wavelengths (e.g., F1130W--F2100W and F1500W--F2100W) find redder sources that may not have detections in shorter wavelength bands such as embedded YSOs which are enshrouded in envelopes too dense for shorter wavelengths to escape. Evolved stars are unlikely to contribute to point source detections in these longest-wavelength filters \citep[e.g.,][]{Jones2015b}. Finally, combinations spanning broad wavelength ranges (e.g., F770W--F2100W or F1000W--F2100W) capture sources apparent across the majority of the mid-IR, likely the brightest and most massive YSOs, or more-evolved YSOs which still retain a disk detectable from the longest wavelength mid-IR emission. 

\begin{table}
\centering
\caption{Table of MIRI color cuts used to identify sources with infrared excess}
\label{tab:MIRI_color_cuts}
\begin{tabular}{lcr}
\hline
\hline
CMD Color  & Color Cut & Source Count   \\
\hline
\hline
F770W--F1000W    & $> 0.221$ & 305   \\
F770W--F1130W   & $> 0.959$ & 457   \\
F770W--F1500W   & $> 0.705$ & 212   \\
F770W--F2100W   & $> 1.398$ & 84   \\
\hline
F1000W--F1130W   & $> 1.287$ & 217 \\
F1000W--F1500W   & $> 0.383$ & 228 \\
F1000W--F2100W   & $> 1.077$ & 89 \\
\hline
F1130W--F1500W   & $> -2.154$ & 278 \\
F1130W--F2100W   & $> -1.460$ & 99 \\
\hline
F1500W--F2100W   & $> -0.806$ & 109 \\
\hline
\hline
\bf{Total Sources} & --- & \bf{833} \\
\hline
\hline

\end{tabular}
\tablecomments{Embedded YSOs in NGC 346 are expected to pass multiple of the above color cuts. Color cuts on the F1130W-F1500W, F1130W-F2100W, F1500W-F2100W are selected conservatively as to not ignore more evolved YSOs which may be visible in the mid-IR, but have falling SEDs between these wavelengths. 
Color cuts are presented in Vega magnitudes.}
\end{table}

\begin{figure*}
\centering
\includegraphics[scale=.65]{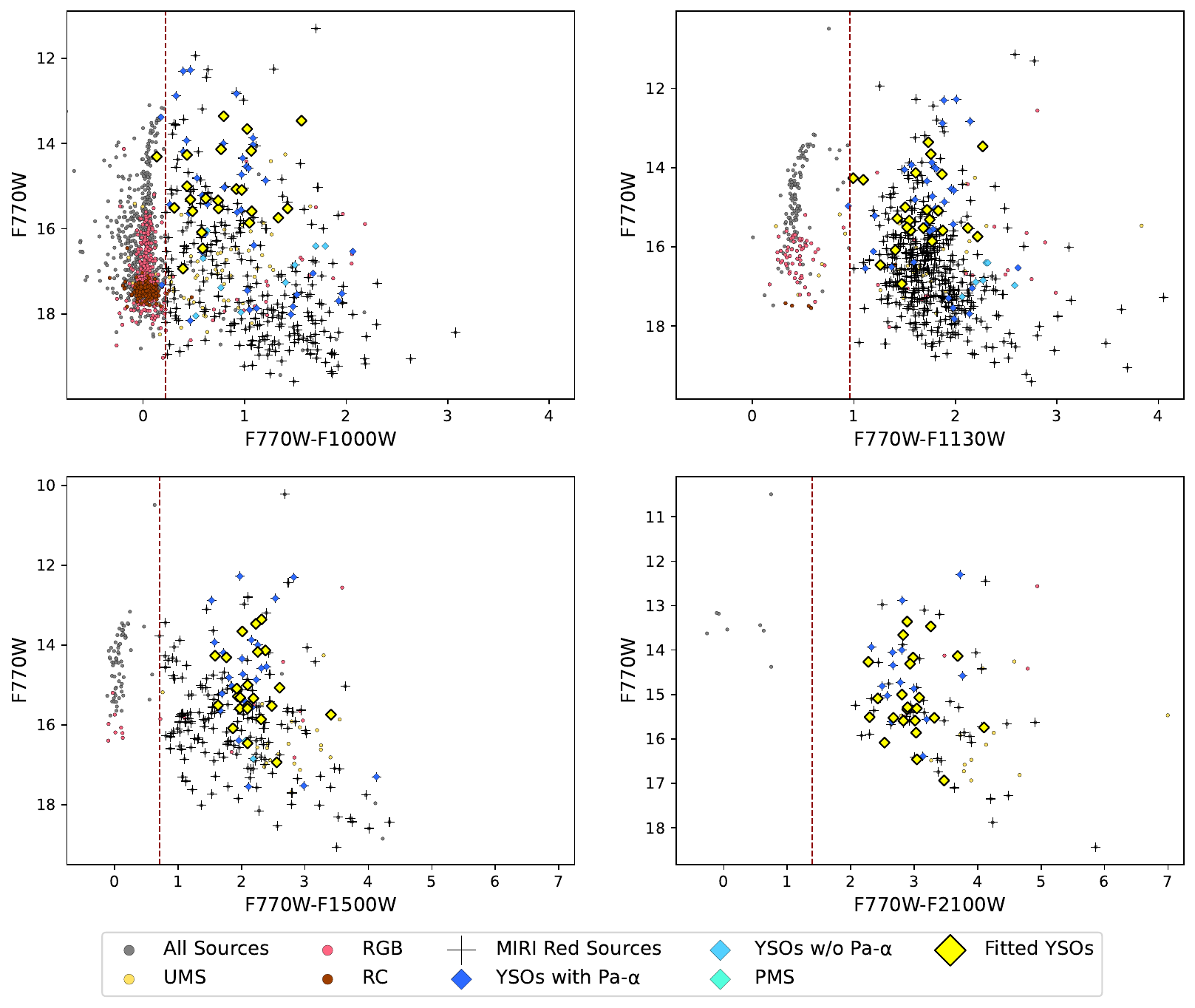}
%\vspace{3cm}
 \caption{ 
CMDSs for MIRI observations. MIRI CMDs show two population groups: sources that are more evolved, consisting mainly of MS, UMS, RGB and RC stars, and sources with IR excess which are likely younger and more embedded. The former, more-evolved population lies near a color of 0, (upper left). 
 The red lines represent a color cut to the right of which detected sources are flagged as candidate YSOs (marked with a black ``+".) 
 Note that a significant population of sources flagged as YSO candidates using MIRI are also identified as YSO or PMS candidates from NIRCam criteria (blue, light blue and teal diamonds). Sources determined to be YSOs based on SED fitting (yellow diamonds) pass color cuts in nearly all cases.} 
 \label{fig:cmd_miri} 
 \end{figure*}

The majority of these MIRI-only CMDs show two populations clearly separated from one another: A tight, vertical population of sources centered at color values of 0 magnitudes, and a more scattered and redder population separated from the vertical population by approximately 1-4 magnitudes in color. \autoref{fig:cmd_miri} shows the four CMDs constructed from the F770W--F1000W, F770W--F1130W, F770W--F1500W and F770W--F2100W colors. The F770W--F1000W $vs$ F770W CMD in particular illustrates the bimodal nature of these population distributions. A significant population of RGB and RC stars are present at a color of $\sim$0, with many NIRCam-identified YSOs separated to the right and redder by up to $\sim$2 magnitudes. For each of our ten CMD combinations, we assign a provisional color cut in order to isolate the redder populations. We use the evolved populations to guide our color cuts, making selections approximately midway between the vertical evolved population and the beginning of the population showing mid-IR excess. Because in this work we are interested in populations of young stars, we remove sources passing these color cuts that have been previously identified as belonging to the RGB, RC, and UMS. In total, 833 objects pass one or more of our ten possible color cuts; we refer to these as ``red" sources and as candidate YSOs. In \autoref{tab:MIRI_RED_table}, we list these sources along with their magnitudes in the MIRI and NIRCam filters where available, and tally the number of color tests each source passes.

\begin{figure*}
\centering
\includegraphics[width=\textwidth]{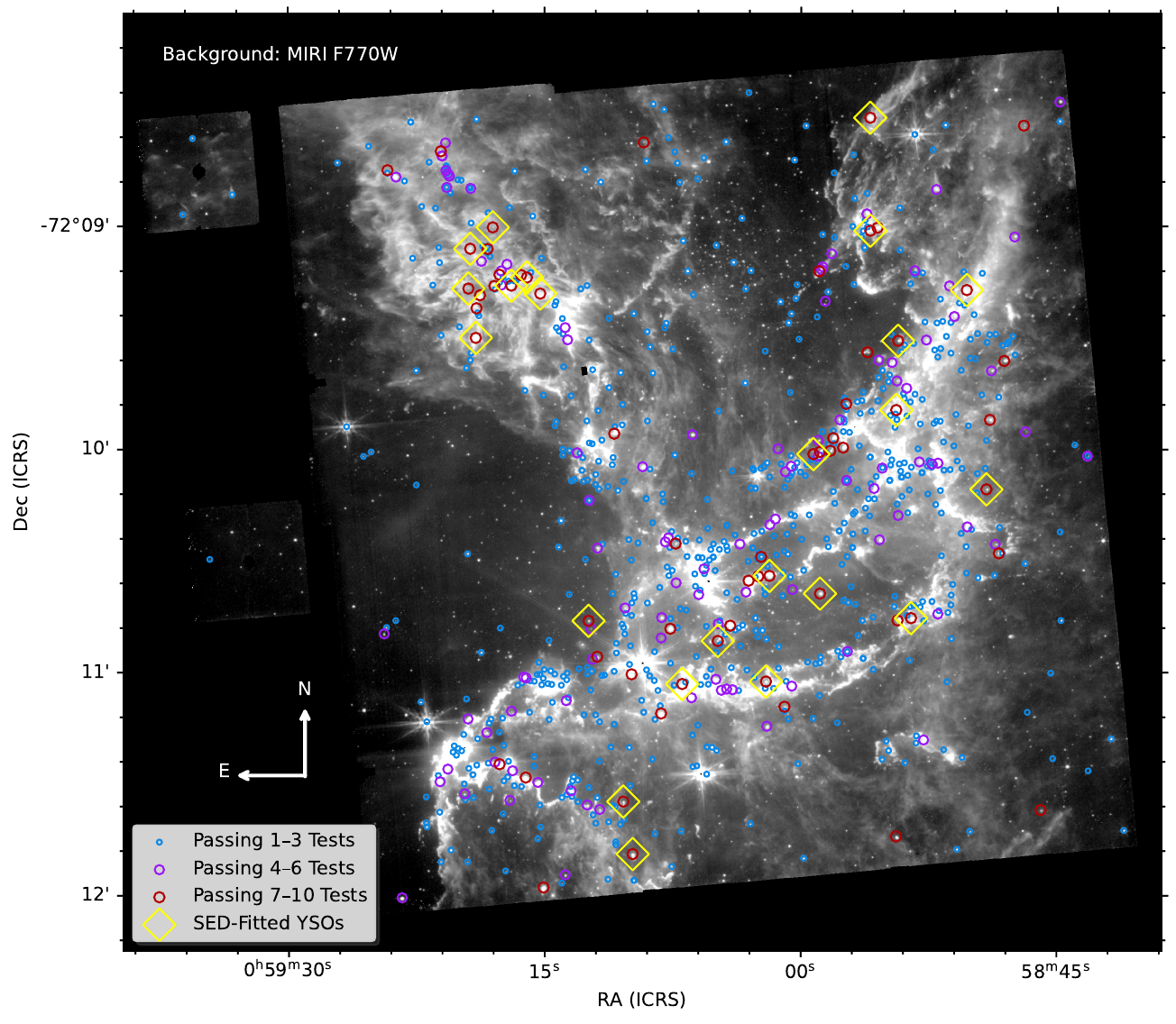}
 \caption{The location in NGC 346 of red, YSOs candidate sources identified with MIRI color criteria. We find a total of 833 red sources using color cuts applied to all 10 possible MIRI color-magnitude diagrams (see \autoref{fig:cmd_miri}). We group these sources into 3 bins based on the number of CMD color tests they pass. Those sources passing more tests are more likely to be young embedded YSOs with rising SEDs. These sources lie more closely along filaments (visible here at 7.7~$\mu$m), indicating that active star formation is occurring along filaments. Those sources passing fewer tests are more dispersed. Yellow diamonds indicate sources containing photometry in all MIRI bands and which have been determined to be likely YSOs based on SED modeling.} 
 \label{fig:MIRI_RED}
 \end{figure*}

Early-stage YSOs still have significant envelopes and thus strong mid-IR contribution, their SEDs display a rising slope in the mid-IR. Thus sources which pass a greater number of color tests, (e.g., are redder in a CMD than the values defined in \autoref{tab:MIRI_color_cuts}), particularly those in the longest mid-IR color combinations, are more likely to be YSOs. In \autoref{fig:MIRI_RED} we plot the spatial distribution of MIRI red sources on our field of view. We group sources together based on the number of color tests they pass. Those passing 1-3 tests are shown with small blue markers, those passing 4-6 with purple markers and those passing 7-10 with larger red markers. We observe a clear correlation of sources along filament structures, particularly among sources passing more color tests.

We observe that many sources detected with NIRCam are also detected in MIRI. When over-plotting the populations identified from their near-IR photometry (such as those described in \autoref{tab:nircam_source_table}) into our MIRI CMDs, we clearly see that the vertical branches are comprised of more-evolved sources along with MS sources as expected. Our F770W--F1000W $vs$ F770W CMD (\autoref{fig:cmd_miri}) in particular contains the most sources in common with NIRCam-identified populations and clearly shows the evolved RGB and RC sources centered at a color of 0. Additionally, we see that many red sources are also those identified as YSOs based on their near-IR color. We note that no PMS stars identified in \autoref{sec:nircam_cmds_ccds} are contained in the mid-IR red population. This is consistent with an isolated star with a disk which has depleted its protostellar envelope and thus does not exhibit significant mid-IR excess. 

\begin{figure*}
\centering
\includegraphics[scale=.6]{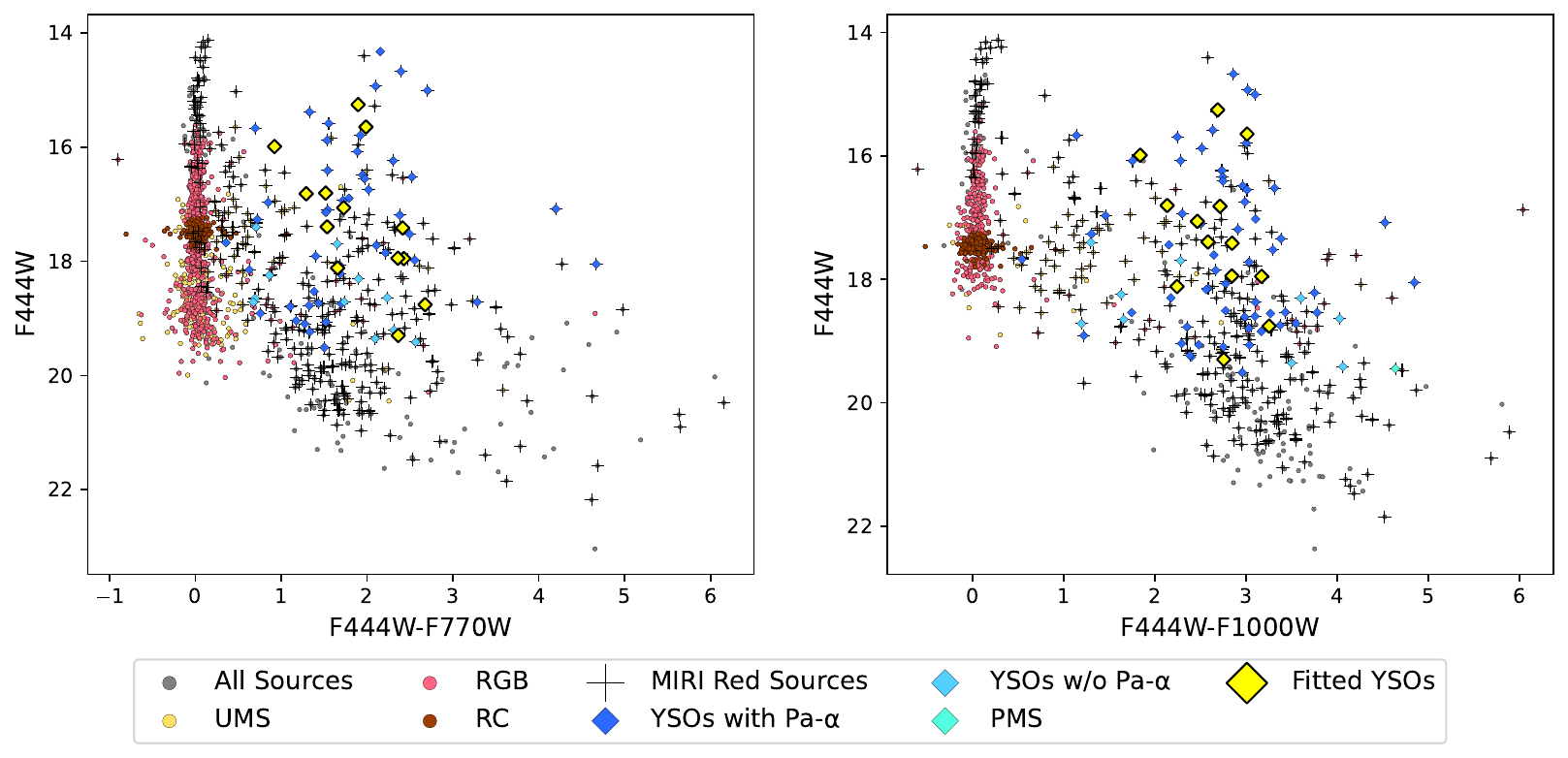}
%\vspace{3cm}
 \caption{
 CMDs combining NIRCam and MIRI filters. Combined photometry from both instruments shows that more-evolved populations identified in the near-IR possess colors near 0 when compared with the mid-IR. Those sources identified as YSOs with NIRCam show significant mid-IR excess, and extend to fainter magnitudes. 
 } 
 \label{fig:cmd_nir_miri}
 \end{figure*}

An examination of CMDs created from both instruments similarly shows a bimodal distribution of populations between the evolved and the redder sources. The F444W-F770W $vs$ F444W and F444W-F1000W $vs$ F444W CMDs (\autoref{fig:cmd_nir_miri}) again show a vertical sequence primarily of RGB and RC stars separated from a population of redder sources extending out $\sim$6 magnitudes and containing many sources identified by NIRCam color analysis as YSOs as well as many sources identified as red from MIRI CMDs. 

We note that in both CMDs we observe a small population of sources classified as ``red" located at the brightest end of the vertical sequence. These sources are not identified as YSOs or RGB stars in NIRCam and are likely those that were bright and saturated in the F115W or F200W band. In a {\em JWST} study of the galaxy NGC 6822, \cite{bib:nally2023} constructed CMDs across these same filters and identified features attributed to supergiants (SGs), oxygen-rich asymptotic giant branch stars (OAGBs), and carbon-rich asymptotoc giant branch stars (CAGBs) located above the tip of the RGB. In their F444W-F770W $vs$ F444W CMD, these populations begin to bend toward redder colors as brightness in the F770W band increases. A similar, though slight, effect can be seen in our CMDs of NGC 346. In CMDs spanning broader colors (e.g., F770W--F1500W $vs$ F770W), \citet{bib:nally2023} showed that the curve towards redder colors of the SG, OAGB and in particular CAGB populations is more pronounced, which is consistent in our case of such sources passing ``red" color cuts in mid-IR CMDs. Therefore, we suggest that these bright sources may belong to these populations of evolved stars.

\newpage

\subsection{YSO SED Fitting}
A total of 77 sources in our catalog have measured photometry in all five MIRI bands (F770W, F1000W, F1130W, F1500W, and F2100W). We further investigate these sources, fitting their photometry to model YSO SEDs. To do this, we use the ``spubhmi” YSO model SEDs from \citet{Robitaille2017} which have been convolved with the NIRCam and MIRI filters by \citet{richardson_2023_8356472}. 
The YSO models span a wide evolutionary range from deeply-embedded, early-stage protostars surrounded by an envelope and accretion disk to late-stage PMS stars whose SED is starting to peak at optical wavelengths. The ``spubhmi" model set includes a central star, passive disk, surrounding envelope, bipolar cavity and ambient medium.

\begin{figure*}
\centering
\includegraphics[scale=.575]{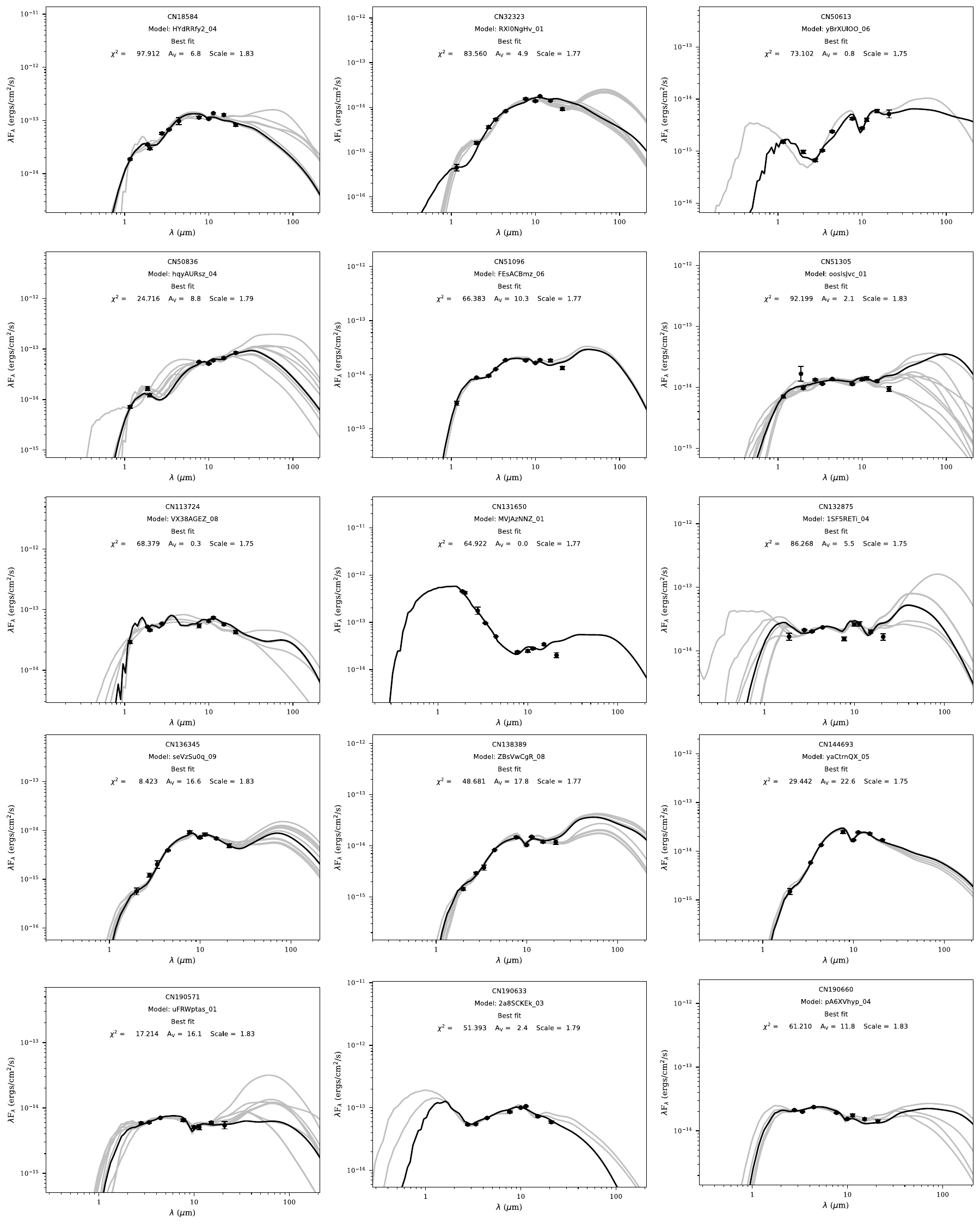}
%\vspace{3cm}
\caption{SED fits for a selection of sources detected in all five filters of our MIRI observations. A total of 23 sources (15 of which are shown here) are identified as YSOs by fitting their photometry to model YSO SEDs. (Continued in \autoref{fig:yso_fits2}.)}
 \label{fig:yso_fits1}
 \end{figure*}

\begin{figure*}
\centering
\includegraphics[scale=.575]{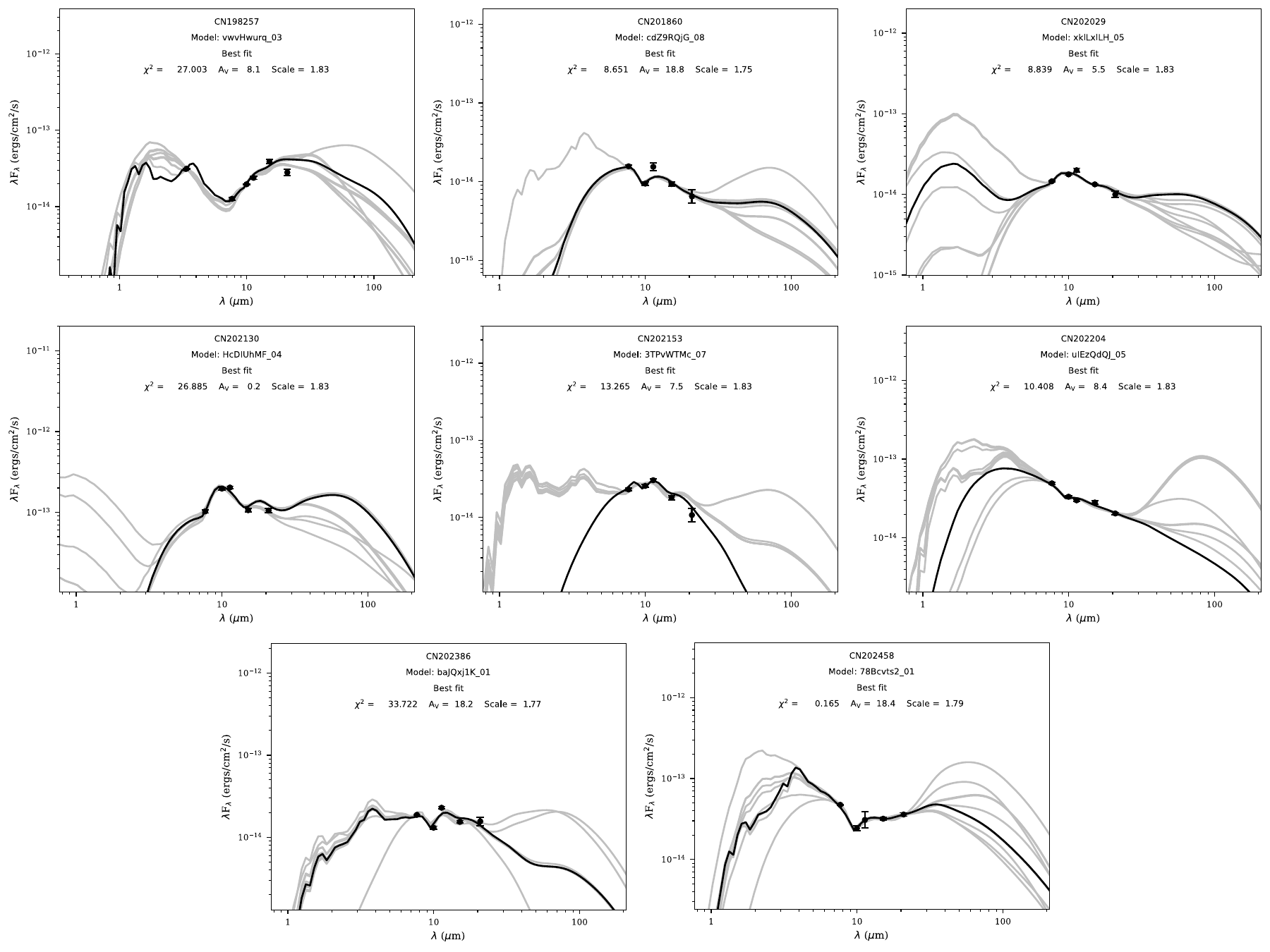}
%\vspace{3cm}
 \caption{SED fits for a selection of YSO-identified sources detected in all five filters of our MIRI observations. (Continued from \autoref{fig:yso_fits1}.)}
 \label{fig:yso_fits2}
 \end{figure*}

In order to determine which of the 77 point sources are YSO candidates, we use a $\mathrm{\chi^2}$ per data point cut of 10, (i.e., the $\mathrm{\chi^2}$ of the best-fit model divided by the number of fitted points is equal to or less than 10). A similar $\mathrm{\chi^2}$ cut was used by \citet{Nayak2023} in the crowded, high-mass, star-forming region 30 Doradus to identify YSO candidates and by \cite{bib:lenkic2023} in the star-forming region Spitzer I in NGC 6822, a nearby barred-irregular galaxy in the local group. Of the 77 sources, 25 meet our $\mathrm{\chi^2}$ per data point cut. Visual inspection confirms that none of these sources show extended morphology consistent with their being background galaxies. However, two such sources are revealed to be small, densely populated clusters (likely of YSOs) when seen at shorter wavelengths. These two sources were thus excluded because of the likelihood of source confusion from the shortest to longest wavelengths, and their MIRI photometry likely containing contributions from multiple sources. We list the model-inferred properties of the remaining 23 YSOs in \autoref{tab:fitted_properties} and show their SEDs in \autoref{fig:yso_fits1} and \autoref{fig:yso_fits2}. We adopt the best-fit model radii and temperatures  and calculate the luminosities and masses with the formulae $\mathrm{L\;=\;4 \pi r^{2} \sigma T^{4}}$ and $\mathrm{L \propto M^{3.5}}$, respectively.  The masses of these YSO candidates ranges from 0.95 -- 4.15 $\mathrm{M_{\odot}}$, showing that we are able to resolve solar-mass YSOs at a distance of 60 kpc with NIRCam and MIRI imaging. We note that the chosen SED models of \citet{Robitaille2017} are calibrated for a galactic gas-to-dust ratio, which has relevance for the inferred envelope infall rate and disk-to-envelope mass ratio. However the best-fit radius and temperature we report here and from which we calculate luminosity and YSO mass, in contrast do not require rescaling for metallicity differences \citep{bib:sewilo2013}. This mirrors the approach of \citet{bib:lenkic2023} and \citet{Nayak2023}.

When comparing these high-confidence SED-fitted YSOs with the larger population of YSOs identified in the near-IR and the ``red" YSO-candidate sources in the mid-IR, we find good agreement as to their nature. Of our 23 fitted sources, two are also classified as YSOs with \paa ~excess based on their NIRCam colors. The remainder lack a detection in one or more of the bands composing the F115W--F187N $vs$ F200W-F444W CCD, and as such are not classified by this criteria. We also compare the 23 fitted sources to the larger sample of 833 objects passing our color cut criteria in MIRI wavelengths. We find that all 23 pass 7 to 10 MIRI color cuts, indicating SEDs rising toward the IR characteristic of young, embedded YSOs. In \autoref{fig:MIRI_RED}, we highlight the locations of these sources with yellow diamonds. 
We note that of these sources, four are also identified as YSOs by {\em Spitzer} surveys of the SMC \citep{Simon2007,bib:sewilo2013}, and two were spectroscopically surveyed in the near-IR with the K-band Multi Object Spectrograph on the Very Large Telescope and shown to be YSOs \citep{bib:jones2022}. Future near-IR spectroscopic studies with JWST of these SED-fitted YSOs and the additional YSO candidates (as determined by NIRCam and MIRI color criteria) will further constrain the nature and properties of these sources. Similar SED analysis of the remaining YSOs which lack a detection in one or more MIRI bands may also yield additional confirmed YSOs, particularly those that are fainter, and possibly less massive, and those at a more-evolved stage with less emission in the mid-IR.

\begin{deluxetable*}{lcccccc}[h]%[p]%[!htbp]
\tiny
\tablecaption{Properties of SED-Fitted YSOs \label{tab:fitted_properties}}
\tablehead{ \colhead{Catalog Number} & \colhead{RA} & \colhead{Dec} & \colhead{Radius} & \colhead{Temperature} & \colhead{Luminosity}  & \colhead{Mass}\\
\colhead{} & \colhead{} & \colhead{} & \colhead{[$\mathrm{R_{\odot}}$]} &  \colhead{[K]} & \colhead{[$\mathrm{L_{\odot}}$]} & \colhead{[$\mathrm{M_{\odot}}$]} }
\decimals

\startdata
\hline
\hline
 CN18584 $^\dagger$ & 14.757664 & -72.176140 &        7.90 &            5139 &     39.2 &      2.85 \\
 CN32323 & 14.704765 & -72.169662 &        0.31 &           16860 &     7.00 &      1.74 \\
 CN50613 & 14.825106 & -72.150103 &        7.38 &            2931 &     3.62 &      1.44 \\
 CN50836 & 14.820602 & -72.154457 &        4.28 &            6415 &    27.94 &      2.59 \\
 CN51096 & 14.816802 & -72.153852 &        3.78 &            7043 &    31.62 &      2.68 \\
 CN51305 $^\dagger$ & 14.813549 & -72.155051 &        2.82 &            6576 &    13.44 &      2.10 \\
CN113724 & 14.733172 & -72.141906 &       43.95 &            2372 &    55.09 &      3.14 \\
CN131650 & 14.745334 & -72.177460 &       18.93 &            4611 &   145.94 &      4.15 \\
CN132875 & 14.778941 & -72.184220 &        5.06 &            6113 &    32.19 &      2.70 \\
CN136345 & 14.709603 & -72.154773 &        1.17 &           14010 &    47.76 &      3.02 \\
CN138389 & 14.726775 & -72.163744 &        2.07 &           12180 &    84.96 &      3.56 \\
CN144693 & 14.733113 & -72.150348 &        2.79 &            8678 &    39.71 &      2.86 \\
CN190571 & 14.831065 & -72.154675 &        4.31 &            7673 &    58.12 &      3.19 \\
CN190633 & 14.829258 & -72.158360 &       17.85 &            3239 &    31.59 &      2.68 \\
CN190660 & 14.830607 & -72.151701 &        7.75 &            6856 &   119.55 &      3.92 \\
CN198257 & 14.801784 & -72.179500 &       36.76 &            2170 &    26.99 &      2.56 \\
CN201860 & 14.726340 & -72.158569 &        0.66 &           15590 &    23.13 &      2.45 \\
CN202029 & 14.746964 & -72.167043 &        5.56 &            6065 &    37.64 &      2.82 \\
CN202130 & 14.791026 & -72.196926 &        0.85 &           17930 &    67.70 &      3.33 \\
CN202153 & 14.793315 & -72.193002 &        0.14 &           14940 &     0.83 &      0.95 \\
CN202204 & 14.758500 & -72.184030 &        1.61 &            9729 &    20.87 &      2.38 \\
CN202386 & 14.770264 & -72.181002 &       28.60 &            2195 &    17.11 &      2.25 \\
CN202458 & 14.723120 & -72.179288 &       82.32 &            2051 &   108.03 &      3.81 \\
\hline
\hline
\enddata
\tablecomments{Column 1: Arbitrary catalog number of sources appearing in this work. Column 2-3: Coordinates. Column 3: Best-fit radius of YSO candidate. Column 4: Best-fit temperature of YSO candidate. Column 5: Luminosity of YSO candidate calculated using the best-fit radius and best-fit temperature with the following formula: $\mathrm{L\;=\;4 \pi r^{2} \sigma T^{4}}$. Column 6: Mass of the YSO candidate calculated using $\mathrm{L \propto M^{3.5}}$. $^\dagger$These sources are also identified as YSOs via NIRCam color criteria. See \autoref{sec:nircam_cmds_ccds}.}
\label{tab:t2}
\end{deluxetable*}

%------------------------------------------------------------------------------------------------
\section{Summary and Conclusions}  
\label{sec:conclusion}

We have observed a low-metallicity star-forming region in the SMC, NGC 346, in the near- and mid-IR with {\em JWST}'s NIRCam and MIRI instruments. We present imaging of this region using both instruments, including the first imaging of this region with MIRI.
This work, based on observations from {\em JWST} GTO program \#1227 (PI:\ M.\ Meixner), has aimed to identify the YSO populations of this region and to characterize the properties of the most confidently detected YSOs, particularly with an eye toward establishing a list of YSOs suitable for upcoming spectroscopic studies. We summarize our findings below.

\begin{itemize}
    \item We image NGC 346, a star-forming region in the low-metallicity SMC, from the near- to the mid-IR across eleven filters using {\em JWST}'s NIRCam and MIRI instruments. This imaging spans a range from 1.15 -- 25.0~$\mu$m.
    \item We observe emission indicating \Hii regions, reservoirs of molecular hydrogen, and filamentary dust structures traced in warm dust and \paa\ emission.
    \item We perform PSF photometry for five of our six NIRCam filter bands, and aperture photometry for the remaining NIRCam band and for our five MIRI filter bands, identifying a total of 203,891 unique sources. 
    \item Using CCDs and CMDs generated from near-IR photometry, we identify young populations, including PMS stars and YSOs often showing \paa ~excess, an indication of ongoing photospheric accretion, identifying $\sim$200 such young sources with a high degree of confidence suitable for upcoming NIRSpec spectroscopic studies. We identify evolved populations such as those belonging to the UMS, RGB and RC.
    \item By identifying a series of color cuts informed by mid-IR CMDs, we identify a population of $833$ objects showing significant mid-IR excess. We show sources in this population are preferentially co-located with regions of diffuse dust and gas, particularly those passing more color cuts, implying many of these objects may be YSOs. A significant number of these objects are also identified as PMS stars or YSOs from their near-IR photometry.
    \item We perform SED fitting on 77 sources detected in all five MIRI bands. Of these 77, 23 are identified as YSOs with a high degree of certainty. From their SED fits, we estimate their radii, bolometric temperatures, luminosities, and masses. These YSOs range in mass from 0.95 to 3.92 $\mathrm{M_{\odot}}$.
\end{itemize}

\vfill\eject

\vspace{0.25 cm}
\section{Acknowledgements}
%\begin{acknowledgments}
This work is based on observations made with the NASA/ESA/CSA James Webb Space Telescope. The data were obtained from the Mikulski Archive for Space Telescopes at the Space Telescope Science Institute, which is operated by the Association of Universities for Research in Astronomy, Inc., under NASA contract NAS 5-03127 for {\em JWST}. These observations are associated with program \#1227.
 Some/all of the data presented in this paper were obtained from the Mikulski Archive for Space Telescopes (MAST) at the Space Telescope Science Institute. The specific observations analyzed can be accessed via~\dataset[10.17909/sz47-c239]{http://dx.doi.org/10.17909/sz47-c239}.

NH and MM acknowledge that a portion of their research was carried out at the Jet Propulsion Laboratory, California Institute of Technology, under a contract with the National Aeronautics and Space Administration (80NM0018D0004).  NH and MM acknowledge support through NASA/{\em JWST} grant 80NSSC22K0025.
OCJ acknowledge support from an STFC Webb fellowship. 
CN acknowledges support from an STFC studentship.
KF acknowledges support through the ESA Research Fellowship.
LL acknowledges support from the NSF through grant 2054178.
ASH is supported in part by an STScI Postdoctoral Fellowship.
ON acknowledges the NASA Postdoctoral Program at NASA Goddard Space Flight Center, administered by Oak Ridge Associated Universities under contract with NASA. Additionally, ON was supported by the director's discretionary fund as a postdoctoral fellow at STScI.  
© 2024 Jet Propulsion Laboratory. All rights reserved.
%\end{acknowledgments}

\vspace{5mm}
\facilities{\em JWST} (NIRCam, MIRI)

\software{1/f removal \citep{bib:1fcor}, 
        astropy \citep{Astropy2013,astropy2018,astropy2022},  
        {\sc topcat} \citep{bib:Taylor2005},
        {\sc starbug ii} \citep{Nally_Starbug2_2023}
        }

%% Appendix material should be preceded with a single \appendix command.
%% There should be a \section command for each appendix. Mark appendix
%% subsections with the same markup you use in the main body of the paper.

%% Each Appendix (indicated with \section) will be lettered A, B, C, etc.
%% The equation counter will reset when it encounters the \appendix
%% command and will number appendix equations (A1), (A2), etc. The
%% Figure and Table counter will not reset.
%\vspace{4cm}
\clearpage
%\vspace{5cm}
\appendix
\section{Additional Tables and Observation Parameters for Primary Observation}

\setcounter{table}{0}
\renewcommand{\thetable}{A\arabic{table}}

\begin{table*}[h]%[p]%[!htbp]
\centering
\caption{NIRCam Observation Parameters for Primary Observation} 
\label{tab:nircam_obs_summary}
\begin{tabular}{ccccccc}
\hline
\hline
Filter & & Read mode & Groups/Int & Integrations/Exp  &  Total Exposure Time (s) & Number of Tiles \\
\hline
\hline
 F115W   & (Mosaic Part 1)           & BRIGHT2 & 2 & 1 &  171.8 & 3 \\
        & (Mosaic Part 2)   & BRIGHT2 & 7 & 1 &  601.3 & 1 \\
\hline
F187N   & (Mosaic Part 1)           & BRIGHT2 & 2 & 1 &  171.8 & 3 \\
        & (Mosaic Part 2)   & BRIGHT2 & 7 & 1 &  601.3 & 1 \\
\hline
F200W   & (Mosaic Part 1)           & BRIGHT2 & 2 & 1 &  171.8 & 3 \\
        & (Mosaic Part 2)   & BRIGHT2 & 7 & 1 &  601.3 & 1 \\
\hline
F277W   & (Mosaic Part 1)           & BRIGHT2 & 2 & 1 &  171.8 & 3 \\
        & (Mosaic Part 2)   & BRIGHT2 & 7 & 1 &  601.3 & 1 \\
\hline
F335M   & (Mosaic Part 1)           & BRIGHT2 & 2 & 1 &  171.8 & 3\\
        & (Mosaic Part 2)   & BRIGHT2 & 7 & 1 &  601.3 & 1\\
\hline
F444W   & (Mosaic Part 1)           & BRIGHT2 & 2 & 1 &  171.8 & 3 \\
        & (Mosaic Part 2)   & BRIGHT2 & 7 & 1 &  601.3 & 1 \\
\hline 
\hline
\end{tabular}
\end{table*}

\vspace{-0.5cm}

\begin{table*}[h]%[!htbp]
\centering
\caption{MIRI Prime Observation Parameters} 
\label{tab:miri_obs_summary}
\begin{tabular}{cccccc}
\hline
\hline
Filter  & Read mode & Groups/Int & Integrations/Exp  &  Total Exposure Time (s) \\
\hline
F770W   & FASTR1 & 28 & 1 &  310.804 \\
F1000W  & FASTR1 & 36 & 1 &  399.606 \\
F1130W  & FASTR1 & 28 & 1 &  310.804 \\
F1500W  & FASTR1 & 32 & 1 &  355.205 \\
F2100W  & FASTR1 & 51 & 1 &  566.108 \\
\hline 
\hline
\end{tabular}
\end{table*}

\vspace{-0.5cm}

\begin{table*}[!h]
    \centering
    \caption{{\sc starbugii} parameters used for aperture and PSF photometry. }
    \begin{tabular}{l|cccccc|ccccc}
    \hline
    \hline
        Parameter &\texttt{F115W} & \texttt{F187N} & \texttt{F200W} & \texttt{F277W} & \texttt{F335M} & \texttt{F444W} & \texttt{F770W} & \texttt{F1000W} & \texttt{F1130W} & \texttt{F1500W} & \texttt{F2100W} \\
        \hline
       \texttt{SIGSRC}          &3.0    &5.0    &3.0    &3.0    &3.0    &3.0        &1.5    &2.5    &2.5    &2.5    &2.5\\ 
        \texttt{SIGSKY}         &1.8    &2.0    &1.8	&1.8	&1.6	&1.8		&1.5	&1.0	&1.0	&1.0	&1.0\\ 
        \texttt{RICKER\_R}      &1.0	&1.0	&1.0	&1.0	&1.0	&1.0		&1.0	&1.0	&1.0	&1.0	&1.0\\ 
        \texttt{SHARP\_LO}      &0.4	&0.4	&0.4	&0.4	&0.4	&0.4		&0.4	&0.4	&0.4	&0.4	&0.4\\ 
        \texttt{SHARP\_HI}      &0.9	&0.9	&0.9	&0.9	&0.9	&0.9		&0.9	&0.9	&0.9	&0.9	&0.9\\ 
        \texttt{ROUND\_LO/HI}   &$\pm1$	&$\pm1$	&$\pm1$	&$\pm1$	&$\pm1$	&$\pm1$     &$\pm1$ &$\pm1$ &$\pm1$ &$\pm1$ &$\pm1$\\ 
        \hline
        \texttt{APPHOT\_R}1.5	&1.5	&1.5	&1.5	&1.5	&1.5	&1.5		&1.5	&1.5	&1.5	&1.5	&1.5\\ 
        \texttt{SKY\_RIN}       &3      &3      &3      &3      &3      &3          &3      &3      &3      &3      &3\\ 
        \texttt{SKY\_ROUT}      &4.5    &4.5    &4.5    &4.5    &4.5    &4.5        &4.5    &4.5    &4.5    &4.5    &4.5\\ 
        \texttt{BOX\_SIZE}      &2      &2      &2      &2      &2      &2          &2      &2      &2      &2      &2\\ 
        \texttt{CRIT\_SEP}      &8      &8      &8      &6      &6      &6          &6      &6      &6      &6      &6\\ 
      \hline
        \texttt{MATCH\_THRESH}  & 0.1 & 0.1 & 0.1 & 0.1 & 0.1 & 0.1 & 0.1 & 0.1  & 0.1 & 0.1 & 0.1\\  
        \texttt{NEXP\_THRESH}   & 3	& 3	& 3	& 3	& 3	& 3	& 3	& 3	& 3	& 3	& 3\\  
        \hline
    \end{tabular}
    \vspace{0.5 cm}
    \label{tab:sb_params}
    %\vspace{5cm}
\end{table*}

\begin{table*}
\centering
\caption{A subset of the 833 sources passing mid-infrared color tests using MIRI aperture photometry. The full table is available in in online format, including associated photometry in NIRCam filters where available. All values are reported in Vega magnitudes.}
\label{tab:MIRI_RED_table}
%\begin{tabular}{lrrllllll}
\begin{tabular}{lcccccccc}
\hline
\hline
Catalog Number &        RA &        Dec & Color Tests Passed &  F770W & F1000W & F1130W & F1500W & F2100W \\
\hline
\hline
         CN72237 & 14.812795 & -72.199420 &          10 &  12.88 &  12.56 &  11.01 &  11.35 &  10.07 \\
         CN14240 & 14.695755 & -72.142483 &          10 &  17.28 &  14.97 &  13.23 &  14.03 &  12.79 \\
         CN16464 & 14.767222 & -72.179861 &          10 &   12.3 &  11.91 &  10.41 &   9.48 &   8.58 \\
        CN202230 & 14.791295 & -72.183496 &           9 &  14.28 &  14.46 &  12.74 &  13.48 &  11.84 \\
        CN202386 & 14.770264 & -72.181002 &           9 &  15.31 &  14.85 &  13.57 &  13.34 &  12.27 \\
        CN132040 & 14.691440 & -72.193612 &           9 &  16.41 &   16.0 &  15.31 &  14.58 &   13.4 \\
        CN201860 & 14.726340 & -72.158569 &           9 &  15.51 &   15.2 &  13.98 &  13.88 &  13.21 \\
         CN18584 & 14.757664 & -72.176140 &           9 &  13.36 &  12.57 &  11.63 &  11.04 &  10.47 \\
         CN50613 & 14.825106 & -72.150103 &           9 &  16.94 &  16.54 &  15.46 &  14.38 &  13.46 \\
        CN202366 & 14.784098 & -72.186422 &           9 &   13.1 &  13.04 &  11.21 &  11.65 &   9.94 \\
         CN50642 & 14.823487 & -72.153640 &           9 &  14.34 &  13.36 &  12.62 &  12.33 &  11.68 \\
        CN150721 & 14.850743 & -72.145830 &           9 &  16.74 &  15.07 &  14.38 &  14.14 &  13.36 \\
        CN132380 & 14.781838 & -72.180093 &           9 &  14.06 &  12.95 &  11.99 &  11.03 &  10.08 \\
         CN50836 & 14.820602 & -72.154457 &           9 &  14.14 &  13.37 &  12.53 &  11.76 &  10.45 \\
        CN132382 & 14.780566 & -72.173728 &           9 &  15.86 &  15.09 &  14.57 &  13.63 &  12.07 \\
         CN51002 & 14.818217 & -72.153668 &           9 &  18.43 &  15.36 &  14.95 &   14.1 &  12.57 \\
         CN51096 & 14.816802 & -72.153852 &           9 &  15.33 &   14.6 &  13.79 &  13.15 &  12.44 \\
         CN88562 & 14.788351 & -72.143760 &           9 &  17.35 &  15.38 &  14.21 &  14.46 &  13.14 \\
         CN51305 & 14.813549 & -72.155051 &           9 &  15.86 &  14.81 &  14.09 &  13.55 &  12.83 \\
        CN144693 & 14.733113 & -72.150348 &           9 &   15.0 &  14.57 &   13.5 &   12.9 &  12.19 \\
        CN132864 & 14.799678 & -72.182181 &           9 &  15.59 &  14.61 &  13.78 &  12.86 &  12.12 \\
        CN132875 & 14.778941 & -72.184220 &           9 &  15.53 &   14.1 &   13.4 &  13.05 &  12.21 \\
        CN199081 & 14.837753 & -72.144419 &           9 &  15.16 &  14.52 &  13.84 &  13.22 &  11.59 \\
        CN138389 & 14.726775 & -72.163744 &           9 &  15.59 &   15.1 &  14.03 &  13.61 &  12.57 \\
        CN136345 & 14.709603 & -72.154773 &           9 &  16.08 &  15.51 &  14.68 &  14.22 &  13.55 \\
         CN20930 & 14.738967 & -72.163287 &           9 &  15.13 &  14.66 &  13.46 &  12.96 &  11.99 \\
         CN20787 & 14.739716 & -72.166558 &           9 &  12.98 &  11.99 &   11.1 &  10.93 &  10.48 \\
         CN20320 & 14.745370 & -72.166873 &           9 &  14.76 &  14.73 &  13.39 &  13.21 &  12.16 \\
        CN133733 & 14.753977 & -72.185915 &           9 &  15.94 &  15.23 &   14.5 &  13.57 &  12.06 \\
         CN20588 & 14.742135 & -72.165849 &           9 &  13.93 &   13.5 &  12.36 &  12.36 &  11.61 \\
\hline
\hline

\end{tabular}
\end{table*}

%% For this sample we use BibTeX plus aasjournals.bst to generate the
%% the bibliography. The sample631.bib file was populated from ADS. To
%% get the citations to show in the compiled file do the following:
%%
%% pdflatex sample631.tex
%% bibtext sample631
%% pdflatex sample631.tex
%% pdflatex sample631.tex
\FloatBarrier
\bibliography{ngc346.bib}{}
\bibliographystyle{aasjournal}

%% This command is needed to show the entire author+affiliation list when
%% the collaboration and author truncation commands are used.  It has to
%% go at the end of the manuscript.
%\allauthors

%% Include this line if you are using the \added, \replaced, \deleted
%% commands to see a summary list of all changes at the end of the article.
%\listofchanges

\end{document}